\documentclass{aa}  
\usepackage{amsmath}
\usepackage{gensymb}
\usepackage{graphicx}
\usepackage{txfonts}
\usepackage{lipsum}
\usepackage{subcaption}         
\usepackage{lscape}             
\usepackage{placeins}           
\usepackage{tabularx}
\usepackage{xcolor}

\usepackage{hyperref}

\newcommand{\noop}[1]{}

\begin{document}
   \title{Multi-vantage SO/PHI and SDO/HMI magnetic\\field observations of active region NOAA 13465.}



   \author{J.~Sinjan\inst{1}\thanks{\hbox{Corresponding author: J. Sinjan} \hbox{\email{sinjan@mps.mpg.de}}}
        \and J.~Hirzberger\inst{1}
        \and G.~Valori\inst{1}
        \and D.~Calchetti\inst{1}
        \and S.~K.~Solanki\inst{1}
        \and L.~A.~Hayes\inst{2}}

   \institute{Max-Planck-Institut f\"ur Sonnensystemforschung, Justus-von-Liebig-Weg 3,
         37077 G\"ottingen, Germany        
         \and
         Astronomy \& Astrophysics Section, School of Cosmic Physics, Dublin Institute for Advanced Studies, \\DIAS Dunsink Observatory, Dublin D15XR2R, Ireland
         }

   

   \date{Received December 30, 20XX}

 
  \abstract
   {Until the launch of Solar Orbiter, long term studies of active region magnetic fields were restricted to a single vantage point: that from Earth. What further restricted these studies were geometrical effects, such as foreshortening and projection, that plague magnetic field observations when the active region was away from the central meridian.}
   {To track an active region, in this case NOAA 13465, with Solar Orbiter during a perihelion pass and use co-observations made by SDO/HMI and SO/PHI-HRT from different vantage points to explore the systematics plaguing observations away from disc centre.} 
   {We use the SHARP (Space-Weather HMI Active Region Patches) data series from SDO/HMI, and create similar datasets with SO/PHI-HRT observations. We compare and contrast the line-of-sight and the three heliographic magnetic field components. Finally we compare the unsigned radial flux, and unsigned totals of the northward and westward magnetic components using the SHARP masking method to select a common area.}
   {We find that the instrument observing closer to the limb infers a greater unsigned radial magnetic flux compared to that observing at disc centre, with a maximum near approximately $45\degree$. The plage displayed more susceptibility to resolution and projection effects than the active region's sole sunspot. In particular the deduced area of the solar plage drove the observed increase in flux near the limb. Thanks to its higher spatial resolution, the SO/PHI-HRT data showed a smaller effect compared to SDO/HMI near the limb, highlighting a dependence on instrument resolution. Finally, the ratio of the two instruments' vertical magnetic flux exhibits a strong positive correlation with the ratio of the cosine of the instruments' viewing angles.}
   {This study lays the groundwork for understanding the impact of viewing geometry on magnetic field inference, a topic that has not been thoroughly explored due to the lack of simultaneous observations from different vantage points. Our results highlight the importance of considering the viewing geometry when interpreting magnetic field observations, and provide a foundation for future studies that aim to use data from different vantage points, such as those from Solar Orbiter and Earth-based instruments.}

   \maketitle

\section{Introduction}

Active regions harbour the most intense magnetic fields found on the surface of the Sun, the photosphere. They typically contain at least one sunspot, seen as a dark feature in white light images, where strong magnetic fields suppress energy transport from below, resulting in a cool region relative to its surroundings \citep{2003A&ARv..11..153S, 2021A&A...656A..92S}. Active regions can last from a few days to months, undergoing several Solar rotations \citep[][]{2015LRSP...12....1V}. They are also the source regions of the most energetic solar events, i.e. solar flares and coronal mass ejections \citep[][]{2011SSRv..159...19F, 2012LRSP....9....3W}.

Long term studies of individual active regions, until recently, have been restricted to a single vantage point along the Sun-Earth line. For long-living active regions, which survive several solar rotations, their evolution is hidden when on the far side of the Sun. Further restricting analysis is a geometrical problem. Due to the spherical shape of the solar surface, several effects plague the inference of the photospheric magnetic field when the active region is near the limb (edge of the visible solar disc). This further reduces the time span where one has reliable measurements of the magnetic field of active regions. 

These include the projection effect, where one detector pixel near the limb spans a much larger area on the solar surface, decreasing the effective spatial resolution. Furthermore, most of the signal loss is due to the cancellation of Stokes $V$ signals due to opposite polarity fields within a resolution element \citep[e.g.][]{2004A&A...417.1125K}. These two effects lead to a much lower magnetic field being inferred with respect to the true value \cite[e.g.][]{2024A&A...683A.134M,2024A&A...690A.341S}. As the magnetic field is mostly organised into radial structures with respect to the solar surface, the polarisation signal near the limb is predominantly via the transverse Zeeman effect (perpendicular to the line of sight or longitudinal component). The transverse Zeeman effect intrinsically has lower signals compared to the longitudinal component, and is more sensitive to noise \citep[e.g.][]{2011A&A...527A..29B}. On top of this, due to the increased path in the solar atmosphere that photons near the limb must travel, more absorption takes places, increasing the height of formation of the spectral line. Because of the decreasing temperature with height in the photosphere (limb darkening), this increases the relative noise levels and because of the decreasing field strength with height, this further reduces the signal strength in Stokes $Q,U$. With highly inclined viewing angles, there are more gradients in the intensity and magnetic field along the line-of-sight. The impact of all these effects on the magnetic field inference has been investigated via simulations \cite[e.g.][]{1998A&A...333..721S,2023ApJ...951...23C,2024A&A...690A.341S}, however there has been no focused study of these impacts on the magnetic field inference using simultaneous observations from vastly different vantage points.

With the launch of the Solar Orbiter mission in 2020 \citep{2020A&A...642A...1M}, carrying a magnetograph on board \cite[Polarimetric and Helioseismic Imager, SO/PHI,][]{2020A&A...642A..11S}, a new vantage point out of the Sun-Earth line has been available for solar magnetic field measurements. At closest approach, Solar Orbiter reaches $0.28$ au, while its aphelion lies near 1 au. The High Resolution Telescope of SO/PHI \cite[HRT,][]{2018SPIE10698E..4NG} enables high resolution magnetic field measurements at close approaches to the Sun from significantly different viewing angles, while the Full disc Telescope (SO/PHI-FDT) observes the entire disc, mostly outside these close approaches. \cite{2025A&A...704A.105K} and \cite{2025A&A...697A.217F} recently exploited far-side observations from SO/PHI-FDT in combination with Earth-based instruments to monitor active regions over multiple solar rotations. 

This work represents a first step towards understanding the impact of viewing geometry on magnetic field inference, a topic that has not been thoroughly explored due to the lack of simultaneous observations from different vantage points. We seek to exploit multi-vantage co-observations of an active region by SO/PHI-HRT and the Helioseismic Magnetic Imager on the Solar Dynamics Observatory \cite[SDO/HMI,][]{2012SoPh..275....3P, 2012SoPh..275..207S, 2012SoPh..275..229S} to directly quantify how inferred magnetic field depend on viewing angles. One of our core aims is to highlight the complementary data from SO/PHI-HRT and SDO/HMI during this co-observation campaign and provide the technical details necessary to reconcile these data sets. We hope to achieve a systematic understanding of the viewing angles effects on the raw magnetic field measurements themselves, as they are used as inputs for many derived quantities of physical importance, such as the magnetic free energy, helicity and coronal magnetic field structure through extrapolations.

We focus on the time evolution of the different components of the magnetic flux of the same target, and attempt to understand any affect of the ever changing viewing geometries by both instruments. In Sect.~\ref{sec:obs} we introduce the observations, morphology and evolution of active region 13465 throughout almost its entire lifetime. In Sect.~\ref{sec:data} we introduce the instruments, data production and additional processing methods. In Sect.~\ref{sec:analysis} we compare maps of the photospheric magnetic field to explore the differences and similarities, as well as the impact of the different viewing directions. In Sect.~\ref{sec:radflux} we compare the unsigned radial fluxes using the same method applied in the SDO/HMI maps. We discuss our findings in Sect.~\ref{sec:disc} and summarise our results in Sect.~\ref{sec:sum}. 

\section{Observations and NOAA 13465 morphology}\label{sec:obs}

\begin{figure}
    \centering
    \includegraphics[width=\linewidth]{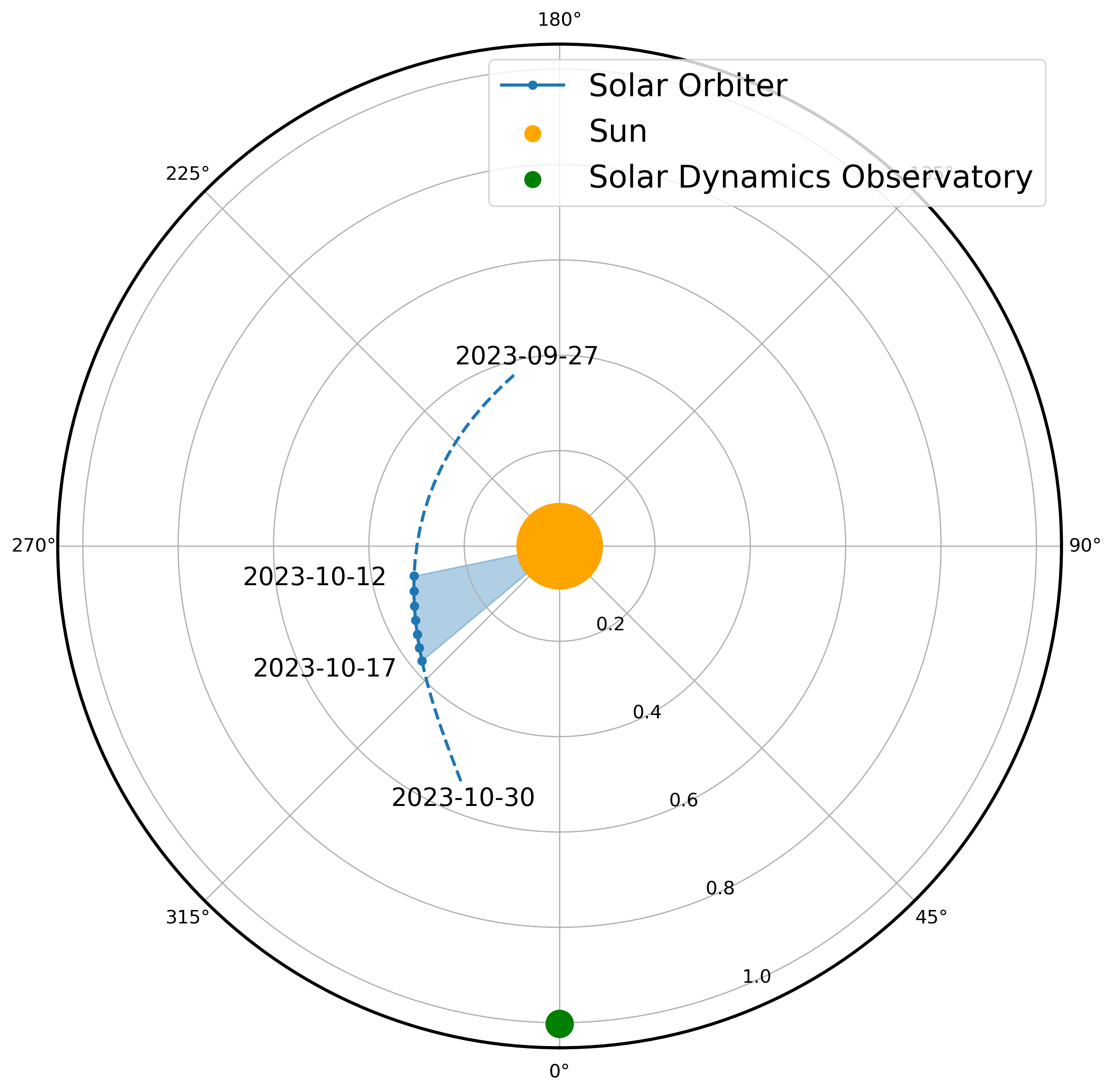}
    {\caption{The orbital trajectory of Solar Orbiter from 2023-09-27 until 2023-10-30 in Geocentric Solar Ecliptic coordinates, with the time spent tracking NOAA 13465 with Solar Orbiter shaded in blue.} \label{fig:orbit}}
\end{figure}

From 12 October 2023 until 20 October 2023 Solar Orbiter conducted a long-term active-region tracking campaign: the \verb|R_SMALL_MRES_MCAD_AR-Long-Term| Solar Orbiter Observing Plan \cite[SOOP, see][for more details]{2020A&A...642A...3Z}. This study tracked an active region with Solar Orbiter's high resolution telescopes as the spacecraft passed Earth quadrature and moved towards the Earth. Figure ~\ref{fig:orbit} illustrates Solar Orbiter's orbit and viewing directions in Geocentric Solar Ecliptic (GSE) coordinates, together with the viewing angle of the active region as seen by SDO/HMI. The longitudinal separation of Solar Orbiter and Earth at the start of observations was $78\degree$, and 8 days later this reduced to $54\degree$. Solar Orbiter was close to the ecliptic plane, with an initial latitudinal separation to Earth of $4.3\degree$, decreasing to $0.4\degree$ 8 days later. This provided a unique opportunity to simultaneously track an active region across the solar disc with Earth-based instruments. 

Using low latency images from SO/PHI-FDT and SO/EUI-FSI a newly emerged active region was spotted on 2\textsuperscript{nd} October on the far side of the Sun as Solar Orbiter was moving towards quadrature \citep[][]{2020JATIS...6d8004A}. These low latency images and those from the following days until the start of the co-observations are displayed in Appendix~\ref{append:fdtll}. This active region emerged at the location of a recently decayed short-lived active region (NOAA 13441) that appeared for only $6$ days during the previous Carrington rotation on the Earth-side. Initially two groups of pores formed, with penumbra forming in both groups a day later. The westward group of pores coalesced into a single large sunspot, while the magnetic fields in the trailing group only grew. Just after the enhancement of the magnetic fields of this trailing group an additional pair of pores formed between the sunspot and trailing group. Over the course of the next $3-4$ days both this trailing group and the newly emerged pair of pores completely decayed, while the umbra in the main sunspot broke up into three distinct groups.

As the selected active region moved into view inside the East limb as seen by Earth, a day prior to the start of the Solar Orbiter campaign, it was assigned the number NOAA 13465. At the start of the campaign at 2023-10-12 00:00 UTC, the sunspot of the active region was located at a longitude of $70\degree$ East as seen from Earth. This corresponded to a $\mu=cos(\theta)$ value of $0.35$, where $\theta$ is the heliocentric angle. As seen from Solar Orbiter in the beginning, the sunspot was very close to disc centre at $\mu=0.98$. Over the next $5.5$ days that this active region was co-observed, it moved towards disc centre as seen from Earth and towards the limb in Solar Orbiter (reaching a minimum of $\mu=0.56$ in the sunspot for Solar Orbiter). After 2023-10-17 10:15 UTC Solar Orbiter ceased observing this active region to carry out another observing campaign. At 2023-10-17 21:00 UTC the active region long-term tracking campaign resumed but Solar Orbiter switched to a different target.

\begin{figure*}
    \centering
    \includegraphics[width=0.88\textwidth]{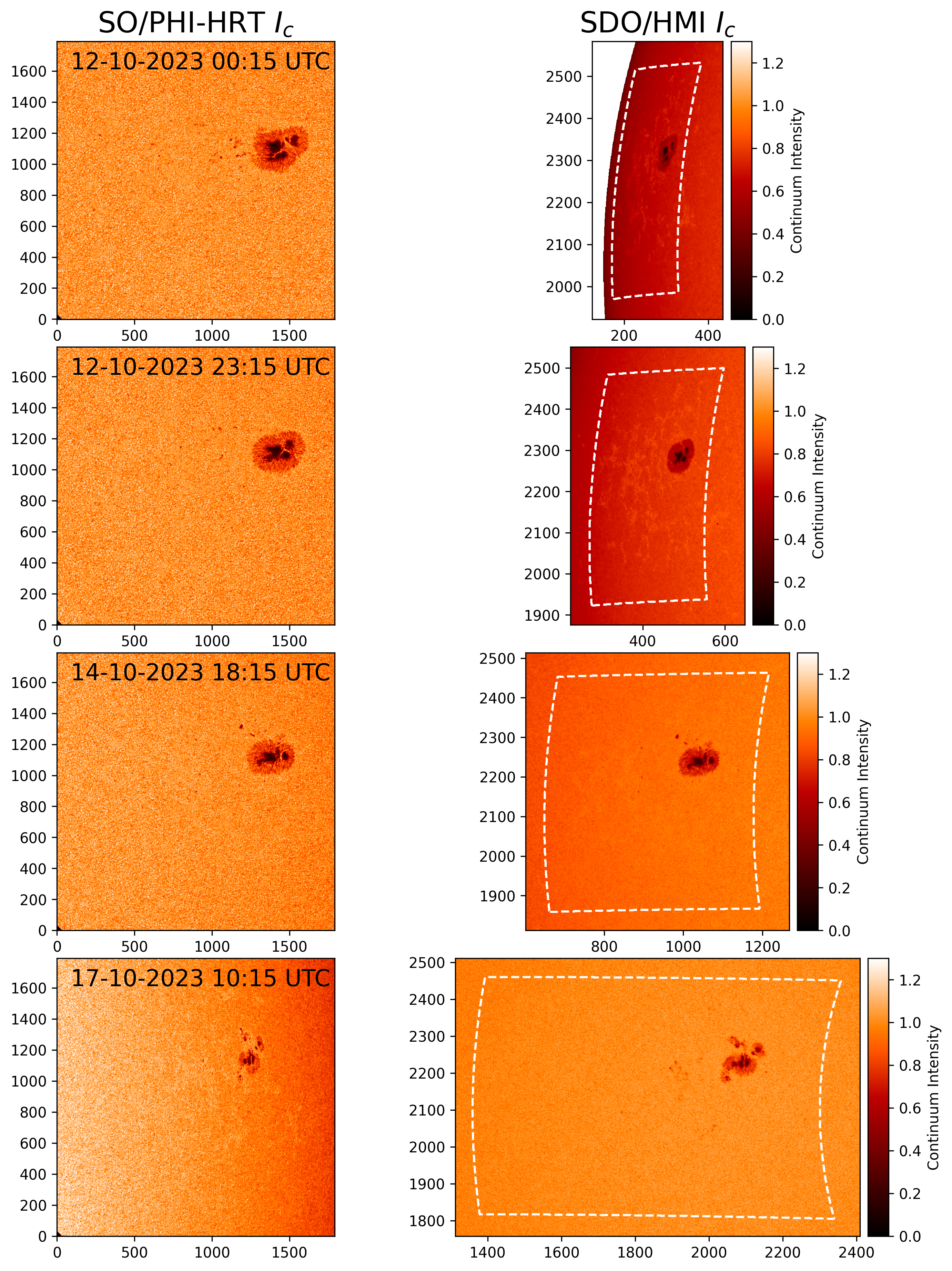}
    {\caption{Normalised continuum intensity images of the active region as co-observed by SO/PHI-HRT and SDO/HMI. First row: start of the co-observations. Second row: the sunspot at $\mu=0.56$ in SDO/HMI and near disc centre in SO/PHI-HRT. Third row: the sunspot at $\mu=0.85$ for both instruments but seen from opposite directions. Final row: the end of the co-observations with the sunspot at $\mu=0.56$ in SO/PHI-HRT and at disc centre in SDO/HMI. Note the difference in spatial resolution and spatial extent. The time of SO/PHI-HRT observations is shown in the text overlay with the nearest $720$s observation from SDO/HMI plotted beside them. The FoV of SO/PHI-HRT is indicated by the dashed white contours in the SDO/HMI cutout. The axes are in units of SO/PHI-HRT (left column) and SDO/HMI (right) pixels.} \label{fig:icnt1}}
\end{figure*}

Fig.~\ref{fig:icnt1} shows continuum intensity images of this active region as seen by SO/PHI-HRT and SDO/HMI during the co-observations. Initially the region was relatively stable, with the distinct umbrae in the leading spot slowly rearranging themselves. After 3 days, on the 15\textsuperscript{th} October, the sunspot started to decay, with pores separating themselves from the sunspot umbrae and the sunspot itself breaking up. Similar figures displaying the line-of-sight magnetic field are shown in Appendix~\ref{append:blos}. After the period of co-observations, the sunspot continued to decay as it moved towards the West limb, as seen by Earth, with the sunspot completely decayed by 22\textsuperscript{nd} October, leaving behind a region of plage. 

\section{Instrumentation and Data}\label{sec:data}

\subsection{SO/PHI-HRT}\label{subsec:hrt}

 SO/PHI measures the polarisation of light from five points in the \ion{Fe}{i}\,$6173.3$\,\AA\ absorption line with an additional point in the nearby continuum. From these polarisation signals SO/PHI infers maps of the photospheric vector magnetic field and line-of-sight velocity via a radiative transfer inversion \cite[MILOS][]{2007A&A...462.1137O}. In addition maps of the continuum intensity are produced. SO/PHI-HRT has a plate scale of $0.5"$ which corresponds to a pixel size of $101\;$km on the surface of the Sun at $0.28$~au. At the start of this campaign Solar Orbiter had just passed perihelion and was at a distance of $0.31$ au, while at the end, after 6 days, the distance increased to $0.36$ au. This resulted in a $3.2$ greater spatial resolution than that of SDO/HMI initially, decreasing to a factor of $2.8$.
 
 SO/PHI-HRT observed at a 1 hour cadence during this long term tracking campaign, with a 3 hour interruption on the morning of 15\textsuperscript{th} October due to mandatory spacecraft reaction wheel offloading maneuvers. These observations were taken in the $60$-second standard operating mode, without any on-board averaging and the Image Stabilisation System \citep[ISS,][]{2012SPIE.8442E..4PV} of SO/PHI-HRT was engaged to track the selected target. The data were reduced using the on-ground pipeline \cite[][]{2022SPIE12189E..1JS, 2023A&A...675A..61K, 2024A&A...681A..58B} with a stray light correction \cite{sinjan2026_underreview}, and are part of the 5\textsuperscript{th} SO/PHI-HRT data release\footnote{\href{https://www.mps.mpg.de/solar-physics/solar-orbiter-phi/data-releases}{5\textsuperscript{th} SO/PHI-HRT data release notes}}. 

Like SDO/HMI, SO/PHI-HRT infers the magnetic field in an observer-based spherical coordinate system, producing the magnetic field strength, inclination with respect to the line-of-sight (LoS), and ambiguous azimuth in the plane of the sky. To produce a disambiguated azimuth we apply the minimum energy method (ME0) \cite[][]{1994SoPh..155..235M,2009SoPh..260...83L}, the same as applied to the SDO/HMI data. However the masking and some of the parameters are adapted to the SO/PHI-HRT data properties: we apply a polarimetric signal threshold to select the pixels where to apply this disambiguation method. For non-selected pixels, the azimuth direction is determined randomly. This implementation is somewhat different to that of SDO/HMI and is described in further detail in Sect.~\ref{subsec:hmi}.

To produce the radial, northward and westward vector components (Heliocentric spherical coordinates) that SDO/HMI provides in the SHARP time series \cite[Space-Weather HMI Active Region Patches,][see~\ref{subsec:hmi} for a brief introduction]{2014SoPh..289.3549B} we follow similar steps to those laid out in \cite{2013arXiv1309.2392S} by first computing the LoS, and two transverse components (in the detector frame). Using the $B0$, $P$ and $L0$ angles, the required rotation matrix is found to transform the vector into the heliocentric spherical coordinate system. The $B0$ angle is the disc centre latitude in Carrington coordinates, the $P$ angle is the angle of the instrument detector with respect to solar North, and the $L0$ angle is the heliographic longitude of the observer. We then spatially degrade the SO/PHI-HRT observations using the theoretical Airy function of SDO/HMI. 

Finally, using a Hann kernel and the adaptive reprojection method described in \cite{2004SoPh..219....3D}, implemented in SunPy 7.0 \citep[][]{2020ApJ...890...68S}, we remap the radial, northward and westward components into the same Cylindrical Equal-Area (CEA) grid as used in the SDO/HMI SHARP time series, downsampling to pixel sizes of $0.03\degree$, in degrees of Heliographic longitude and latitude on the solar surface. This pixel size is approximately the size of one SDO/HMI pixel ($0.5\arcsec$) at disc centre observing at 1 au. To align the maps from SO/PHI-HRT and SDO/HMI we used co-temporal magnetograms from SDO/HMI to account for any pointing errors in the WCS information of the SO/PHI-HRT datasets, and after remapping use the radial magnetic field to correct residual shifts. 

\subsection{SDO/HMI}\label{subsec:hmi}

SDO/HMI also samples the \ion{Fe}{i}\,$6173.3$\,\AA\ absorption line, with six points in the line, but no point in the continuum. It also has a plate scale of $0.5"$ but observes at a distance of 1 au. To produce a continuum intensity map, SDO/HMI reconstructs the spectral line from Doppler-shift, line-width, and line-depth estimates \cite[see Eq.~14 of][]{2016SoPh..291.1887C}. Further details comparing the SDO/HMI and SO/PHI-HRT instrument properties can be found in Table~1 of \cite{2023A&A...673A..31S}.

The SDO/HMI SHARP data series from SDO/HMI \cite[][]{2014SoPh..289.3549B} contains patches of automatically tracked active regions on a 12 minute cadence and includes the vector magnetic field. These tracked active regions are denoted with distinct HARP (HMI Active Region Patches) numbers. In this paper we use the definitive SHARP CEA series, which are computed after the AR has completed its passage across the disc. The SHARP CEA series provides the radial, northward and westward magnetic field components on a CEA grid. To disambiguate the azimuth such that the vector can be transformed, the SHARP CEA series use a transverse magnetic field strength threshold, that is dependent on the position on the solar disc, to determine pixels with a high confidence of disambiguation and there apply the ME0 method. Furthermore those pixels separated by no more than 5 pixels from these high confidence pixels (denoted as intermediate confidence pixels) are also disambiguated with ME0 but with smoothing applied. For all other pixels, the radial-acute-angle solution is used.

NOAA 13465 has a HARP number of 10248. The SHARP time series on the CEA grid is produced at a 720 second (12 minute) cadence for the duration of the HARP as it moves across the solar disc as seen by SDO/HMI and uses the standard 720 second vector magnetic field to compute the radial, westward and southward components \cite[][]{2014SoPh..289.3483H}. As explained in \cite{2013arXiv1309.2392S}, the southward component $B_T$ provided in the SHARP data series is equivalent to $-B_y$ of the Heliographic components described in \cite{1990SoPh..126...21G}. We therefore multiplied the SDO/HMI $B_T$ by -1 to align with the northward direction of $B_T$ from SO/PHI-HRT. The SHARP magnetograms are derived from the non-vector, i.e. purely LoS, \verb|m_720s| data series. In this work we have applied no post-processing to the SDO/HMI SHARP datasets. The first SHARP observations of HARP 10248 begins at 2023-10-10 15:24 TAI (International Atomic Time), when only a very small fraction of the active region is visible as seen by SDO. 

\subsection{Cross-calibration}\label{ssec:cc}

From \cite{sinjan2026_underreview} a good agreement was found between the stray light corrected SO/PHI-HRT and standard SDO/HMI data in the three vector magnetic field components. Only for ($|\mathbf{B}|>1.6\;$kG) were differences found: SDO/HMI inferred field strengths approximately $4\%$ greater than SO/PHI-HRT that were also $10\%$ more vertically inclined. Conversely because the SDO/HMI magnetograms are not computed via a RTE inversion, for the same strong pixels the line-of-sight magnetic field from SDO/HMI is $13\%$ weaker than SO/PHI-HRT. These weaker fields were also found when SDO/HMI magnetograms were compared with their vector counterparts in \cite{2014SoPh..289.3483H}. This comparison was made using data from a one hour period near 12:00 UTC, and given the well-documented, orbit-induced systematics in the SDO/HMI data (at 6, 12 and 24 hour periodicities) we expect the differences between SO/PHI-HRT and SDO/HMI to vary, however this will not impact any observed trends. For this reason we do not apply any cross-calibration factor to the SO/PHI-HRT or SDO/HMI data used in this study.

\section{Photospheric magnetic field}\label{sec:analysis}

For the analysis performed in this section, we include SDO/HMI SHARP data at the nearest time to the SO/PHI-HRT observations at the same hourly cadence. The SO/PHI-HRT observations were at XX:15 UTC, and given the difference in light travel time the nearest SDO/HMI observations were at XX:24 TAI. This resulted in a real time difference of 2-3 minutes, with the time difference decreasing towards the end of co-observations as Solar Orbiter moved closer to the Earth. 

\begin{figure*}
    \centering
    \includegraphics[width=0.92\textwidth]{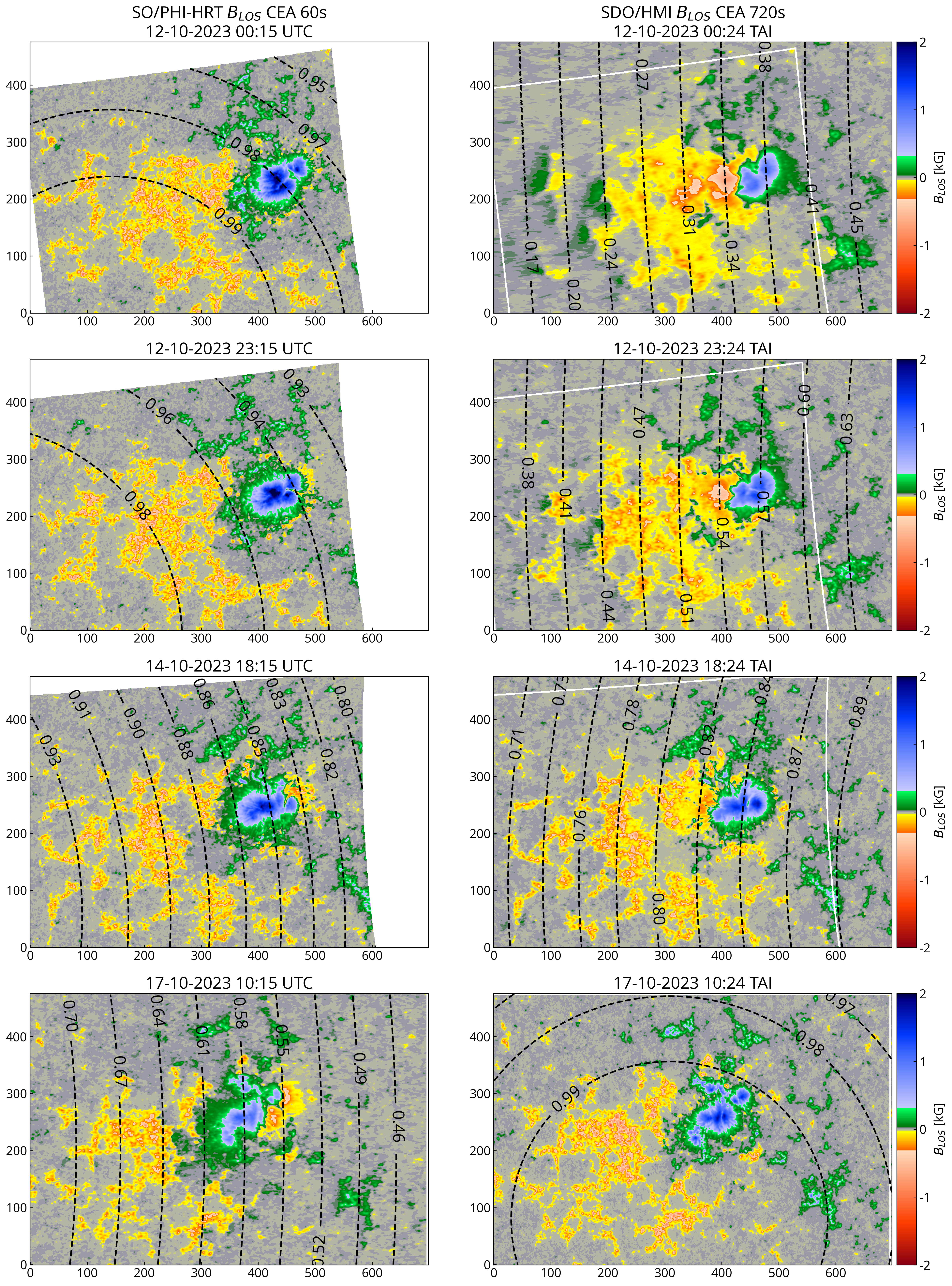}
    {\caption{Line-of-sight magnetic field as inferred by SO/PHI-HRT (left column) and SDO/HMI (right column) in cylindrical equal area coordinates. The four rows are observations at the same times as the rows in Fig.~\ref{fig:icnt1}. The colour map is saturated at $\pm2\;$kG and values below $\pm15\;$G are shades of grey. The contours of constant $\mu$ are overlaid in dashed black lines while the SO/PHI-HRT FoV in white is indicated on the SDO/HMI maps. The axes are in units of SHARP pixels.} \label{fig:blos_4stage}}
\end{figure*}

\begin{figure*}
    \centering
    \includegraphics[width=0.9\textwidth]{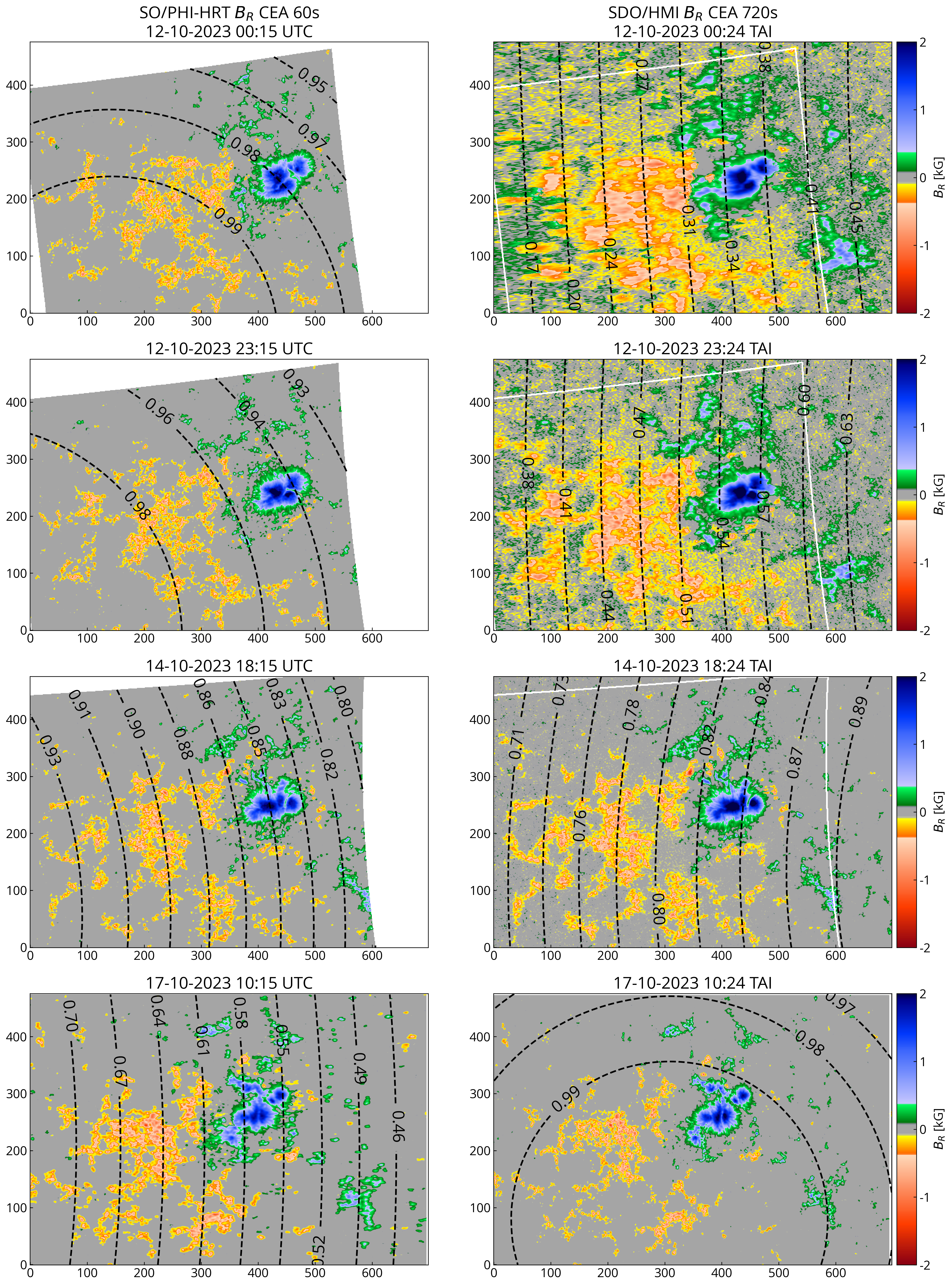}
    {\caption{Radial magnetic field as inferred by SO/PHI-HRT and SDO/HMI in cylindrical equal area coordinates. The four rows are observations at the same times as the rows in Fig.~\ref{fig:icnt1}. The colour map is saturated at $\pm2\;$kG and values below $\pm45\;$G are masked in gray. The contours of constant $\mu$ are overlaid in dashed black lines while the SO/PHI-HRT FoV in white is indicated on the SDO/HMI maps. The axes are in units of SHARP pixels.} \label{fig:br_4stage}} 
\end{figure*}

\begin{figure}
    \centering
    \includegraphics[width=\linewidth]{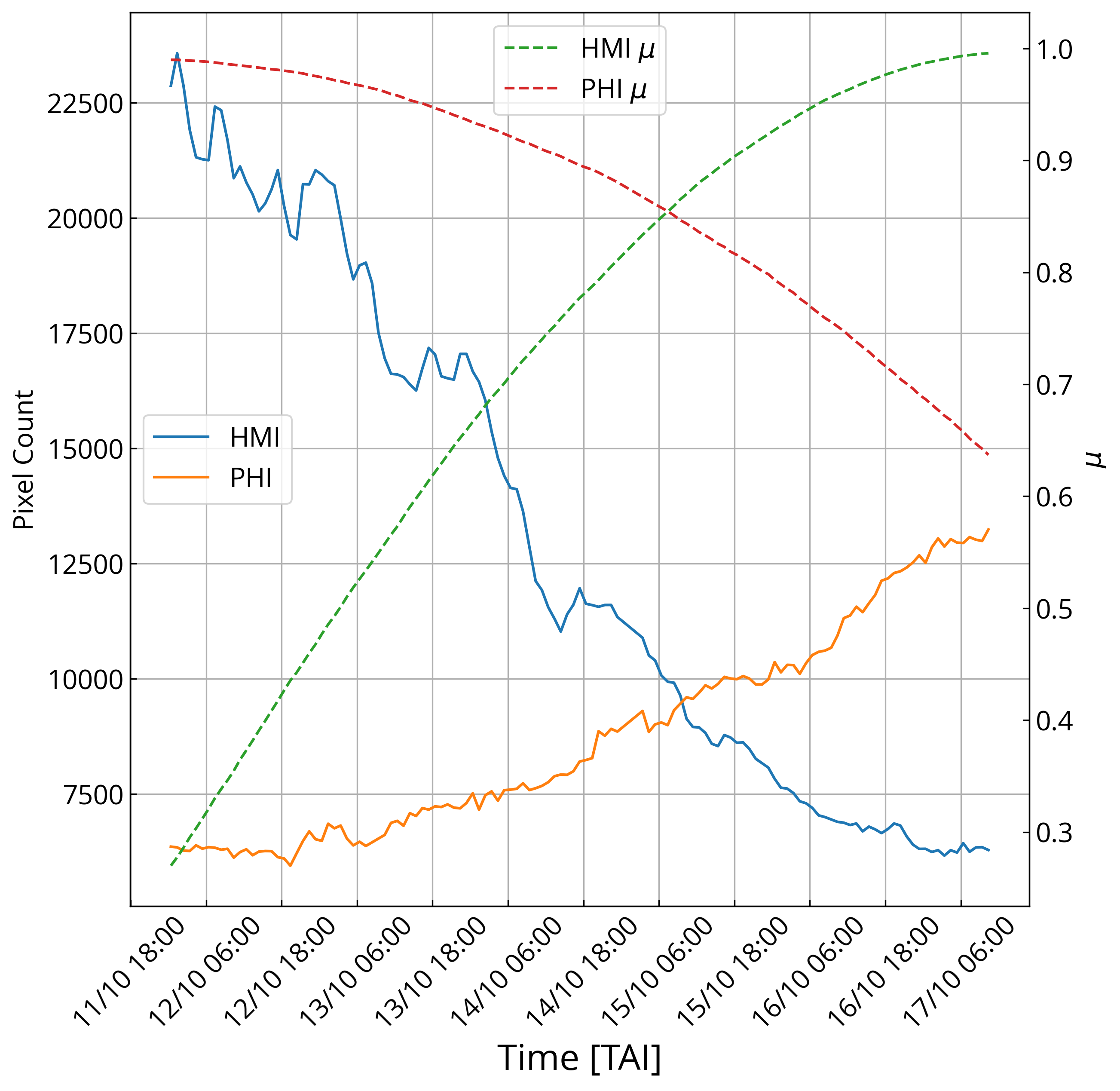}
    {\caption{Pixel count of the negative polarity plage region, using $B_{R}=-300\;$G to define the edge of the plage features with a minimum feature size of $4$ pixels, as seen by both SDO/HMI and SO/PHI-HRT.} \label{fig:plage_area}}
\end{figure}

\begin{figure*}
    \centering
    \includegraphics[width=0.90\textwidth]{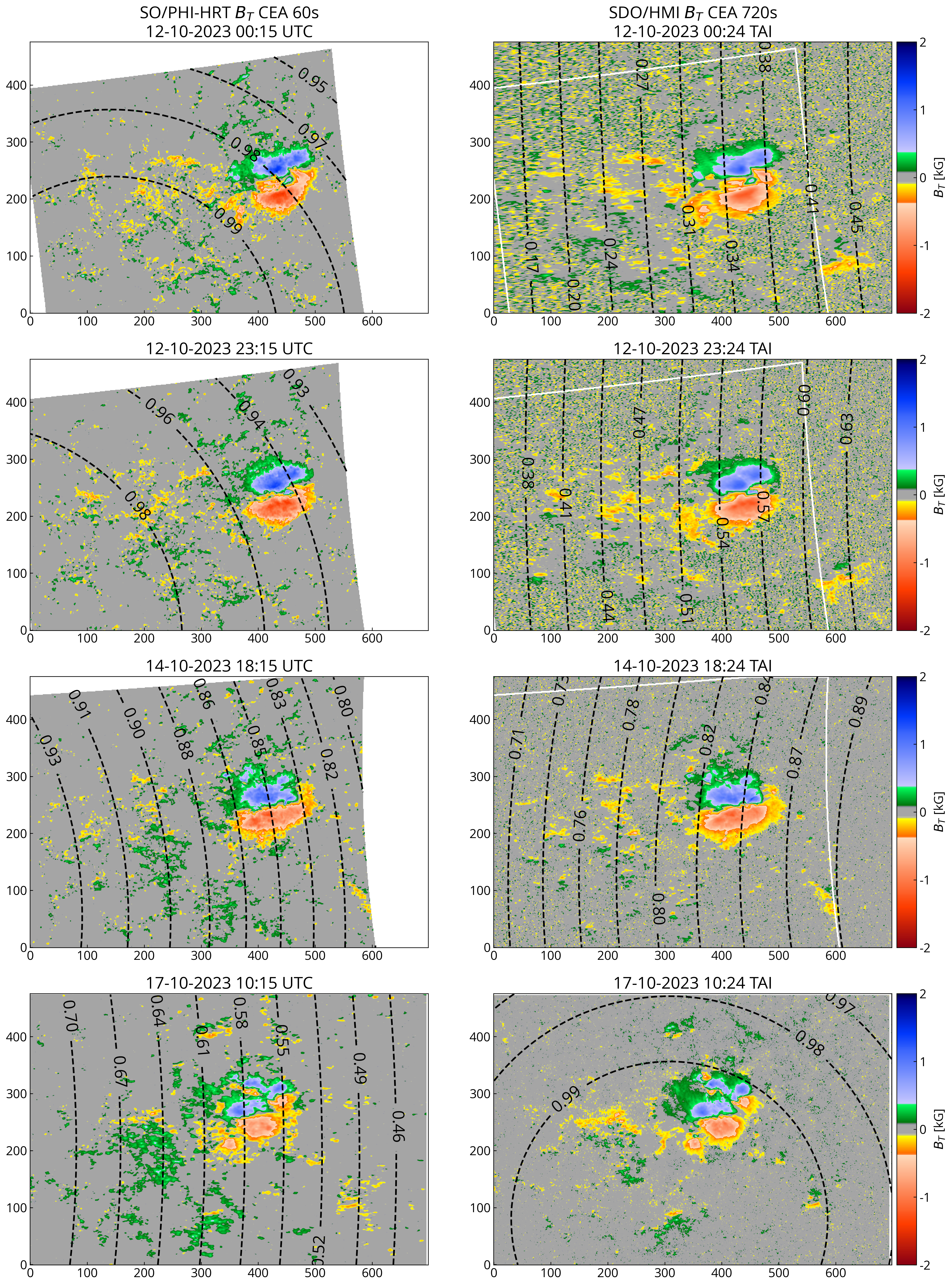}
    {\caption{Northward magnetic field as inferred by SO/PHI-HRT and SDO/HMI in cylindrical equal area coordinates. The four rows are observations at the same times as the rows in Fig.~\ref{fig:icnt1}. The colour map is saturated at $\pm2\;$kG and values below $\pm45\;$G are masked in gray. The contours of constant $\mu$ are overlaid in dashed black lines while the SO/PHI-HRT FoV in white is indicated on the SDO/HMI maps. The axes are in units of SHARP pixels.} \label{fig:bt_4stage}}
\end{figure*}

\subsection{Line-of-sight magnetic field comparison}\label{ssec:los}

In Fig.~\ref{fig:blos_4stage} the reprojected $B_{LOS}$ maps are shown at 4 stages in time during the co-observations, saturated at $\pm2\;$kG. Contours of $\mu$ are overlaid in dashed black lines. The FoV of SO/PHI-HRT does not initially cover the entire region of the SHARP maps, and is outlined by the solid white line in the SDO/HMI maps. The first row corresponds to the first pair of observations, while the bottom row is the final pair of observations (where $\mu_{PHI}=0.56$, $\mu_{HMI}=0.99$ at the sunspot). The second row (where $\mu_{PHI}=0.94$, $\mu_{HMI}=0.56$ at the sunspot) corresponds to almost the reverse of the viewing angle combination of the final observation pair. The third row corresponds to where $\mu_{PHI}$ is almost identical to $\mu_{HMI}$, near a value of $0.85$, at the centre of the sunspot but from opposite directions. 

There are stark differences, as expected, in the polarity and structure of the magnetic features due to the vastly different lines of sight. In the first row of Fig.~\ref{fig:blos_4stage}, where the difference in viewing angle or $\mu$ is largest, one first notices that an opposite line-of-sight magnetic field polarity is found to the limb-side of certain plage features, particularly those closest to the limb (left side of the SDO/HMI map). This plage region, has a clear negative polarity as seen by SO/PHI-HRT, with fields on the order of $0.1-1\;$kG. The opposite polarity regions in SDO/HMI, displayed in dark green, are mostly rather weak fields, ranging from $10-80\;$G. Please note that all field values are those determined at the resolution of the instruments, which may not correspond to the true field values, as the finite resolution of the instruments is not accounted for when deriving the magnetic field from the Stokes parameters. This polarity reversal on the limb-side of plage was recently reported by \citep[][]{2025arXiv251000320A} when analysing multi-vantage SDO/HMI and SO/PHI-HRT data, and was interpreted as being a result of the overlying magnetic canopy following \cite{2010A&A...518A..50P}.

Equally, there are large differences in the sunspot, with SDO/HMI clearly probing the expansion of the magnetic field in higher photospheric layers, with the limb-side area of the sunspot reversing polarity with respect to the disc centre view by SO/PHI-HRT, a well known effect. The observed difference in spatial extent of the sunspot and neighbouring plage plays a large role in the identification of pixels that belong to the active region. The $B_{LOS}$ maps are used for automatic tracking and generation of the \verb|bitmap| segment from the SHARP data series \citep[][]{2010SoPh..262..277T}. The \verb|bitmap| appears not to capture the entire active region initially, when it is very near the limb as seen by SDO/HMI. The \verb|bitmap| segment is used in combination with the \verb|conf_disambig| segment, to select the high confidence pixels used for computing space-weather-relevant quantities, such as the the total unsigned flux, vertical current density etc. \citep[][]{2014SoPh..289.3549B}. The systematics of the SHARP pixel selection procedure near the limb are not the focus of this work, and are reserved for future studies.

As time progresses, and the $\mu$ values of the instruments reverse, so do the differences, such that the two instruments show consistent behaviour when considering the whole time series. Thus, the polarity of the limb side plage region, mentioned above, also reverses in the SO/PHI-HRT data, where there is quiet Sun to the limb-side of the plage, with similar orders of magnitude ($10-50$\;G) in the line-of-sight magnetic field to that found in the SDO/HMI map at the start of the observations. 

\subsection{Radial magnetic field comparison}\label{ssec:br}

In Fig.~\ref{fig:br_4stage} we compare the maps of the radial magnetic field, with respect to the solar surface, during the co-observations. Even with a $\pm45$\;G gray noise mask, one immediately notices the difference in the quiet Sun, with an apparent higher level of background noise in the SDO/HMI maps. This arises from the different disambiguation methods applied in the quiet Sun regions, described in Sect.~\ref{subsec:hrt} and~\ref{subsec:hmi}. On a broad level, however, the properties of the magnetic structures, i.e. their polarity, size and spatial distribution agree well. 

Nevertheless, from the first row of Fig.~\ref{fig:br_4stage} it is clear that the magnetic plage, both the negative polarity region to the left of the sunspot, and the positive polarity region near and above the sunspot disagree in the spatial extent of the structures. Here SDO/HMI infers a greater spatial extension of the plage features, with several distinct features seen by SO/PHI-HRT appearing as one contiguous magnetic feature in SDO/HMI. This discrepancy in plage feature size decreases as both instruments start to observe at similar $\mu$ values, namely the third row in Fig.~\ref{fig:br_4stage}. It is a common effect to both instruments, as the last row shows the same behaviour in the SO/PHI data when it is observing near the limb, although not so strongly. 

This is demonstrated in Fig.~\ref{fig:plage_area}, where the area of all negative polarity plage is compared using $B_{R}=-300\;$G to define the boundary of the plage features, and applying a minimum area of each plage feature of 4 pixels. Initially the area is much greater in SDO/HMI, reaching equivalence with SO/PHI-HRT near the point where the $\mu$ values intersect (indicated by the dotted green and red lines). Following which the area, that we denote here with $A$, is larger in SO/PHI-HRT when it observes near the limb. Nevertheless, the ratio of $A^{PHI}_{\text{plage}}/A^{HMI}_{\text{plage}}$ at the final time step (mean $\mu_{PHI}/\mu_{HMI}=1.563$) is 2.11, while the ratio $A^{HMI}_{\text{plage}}/A^{PHI}_{\text{plage}}$ is 2.33, when the mean $\mu_{HMI}/\mu_{PHI}=1.563$. These area ratio quantities were computed with the curves shown Fig.~\ref{fig:plage_area}, but smoothed with a Gaussian filter with a 6 hour $\sigma$ to suppress the orbital fluctuation in the SDO/HMI data. This difference in the area ratios ($10\%$), when the viewing angle combination is identical but reversed, may be a reflection of the increased spatial resolution of SO/PHI-HRT, minimising projection effects near the limb.

\subsection{Northward and Westward magnetic field comparison}\label{ssec:btbp}

In Fig.~\ref{fig:bt_4stage}, the northward components of the magnetic field are compared. From the four panels, there is a good agreement in and near the sunspot. Away from the two large polarity regions in the sunspot, it is clear that there is a higher background noise in the SDO/HMI maps, most evident in the first two panels. This is a result of the different disambiguation methods in this region. In the neighbouring plage regions, the agreement is less clear. As plage regions are predominantly vertical we do expect very minor horizontal components. 

\begin{figure*}
    \centering
    \includegraphics[width=0.90\textwidth]{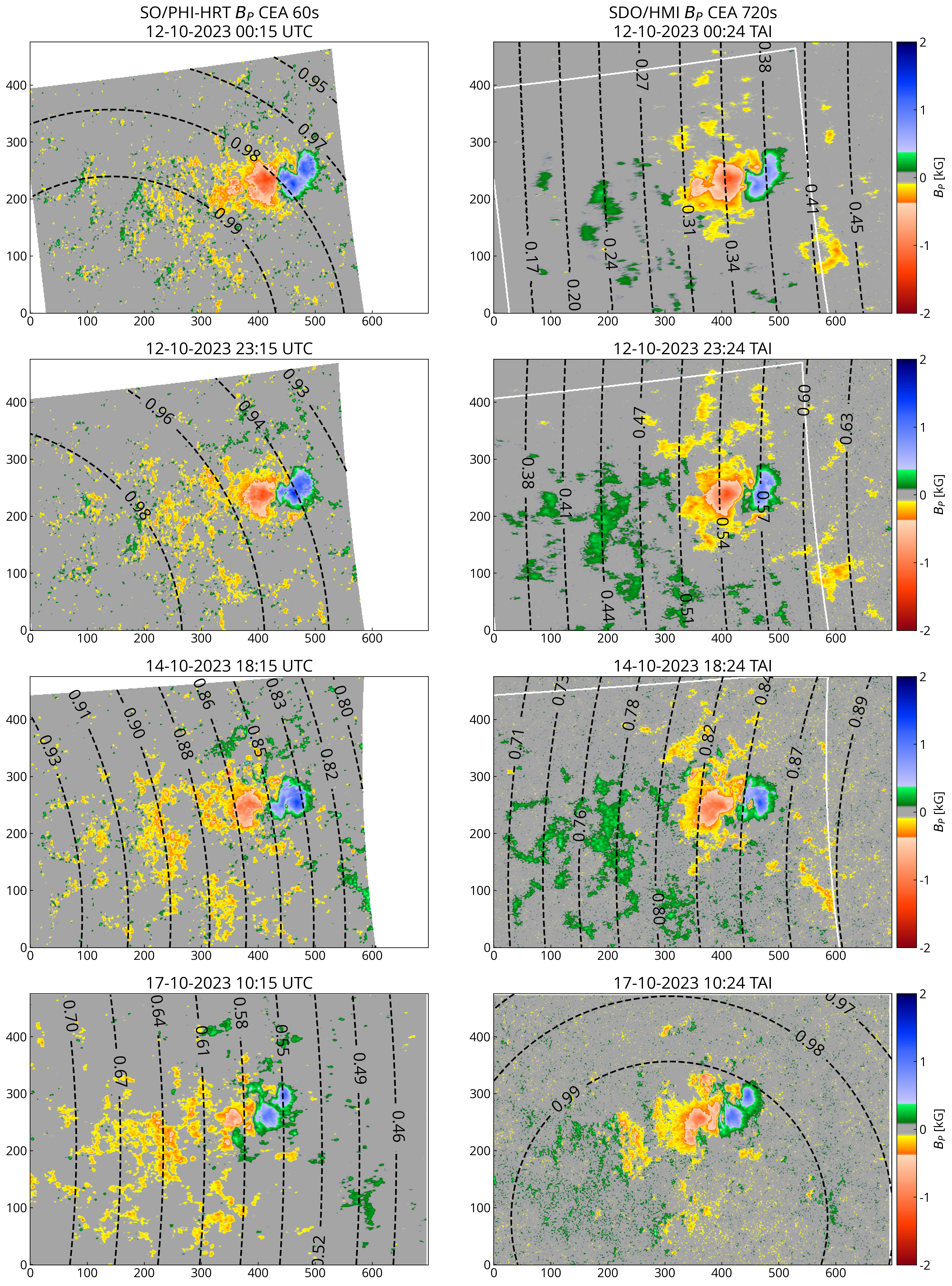}
    {\caption{Westward magnetic field as inferred by SO/PHI-HRT and SDO/HMI in cylindrical equal area coordinates. The four rows are observations at the same times as the rows in Fig.~\ref{fig:icnt1}. The colour map is saturated at $\pm2\;$kG and values below $\pm45\;$G are masked in gray. The contours of constant $\mu$ are overlaid in dashed black lines while the SO/PHI-HRT FoV in white is indicated on the SDO/HMI maps. The axes are in units of SHARP pixels.} \label{fig:bp_4stage}}
\end{figure*}

A similar picture presents itself when comparing the westward components in the sunspot, depicted in Fig.~\ref{fig:bp_4stage}: with components of similar magnitude, distribution and spatial extent. However, the polarity of the plage region to the left of the sunspot displays a contradiction: the polarities of the westward component disagree, with an eastward (yellow) polarity inferred by SO/PHI-HRT and a westward (green) polarity inferred by SDO/HMI. In the lower panel the polarity inferred by SDO/HMI appears to flip when this plage region cross the central meridian. This is most likely due to the reported East-West hemispheric bias in the SDO/HMI data \citep[][]{2022SoPh..297...17L}. SO/PHI-FDT also exhibits the same East-West hemispheric bias, however it has not been possible to confirm if the same behaviour exists in SO/PHI-HRT. 

\section{Vertical magnetic flux and horizontal components comparison}\label{sec:radflux}
\subsection{Vertical magnetic flux}

\begin{figure*}
    \centering
    \includegraphics[width=\textwidth]{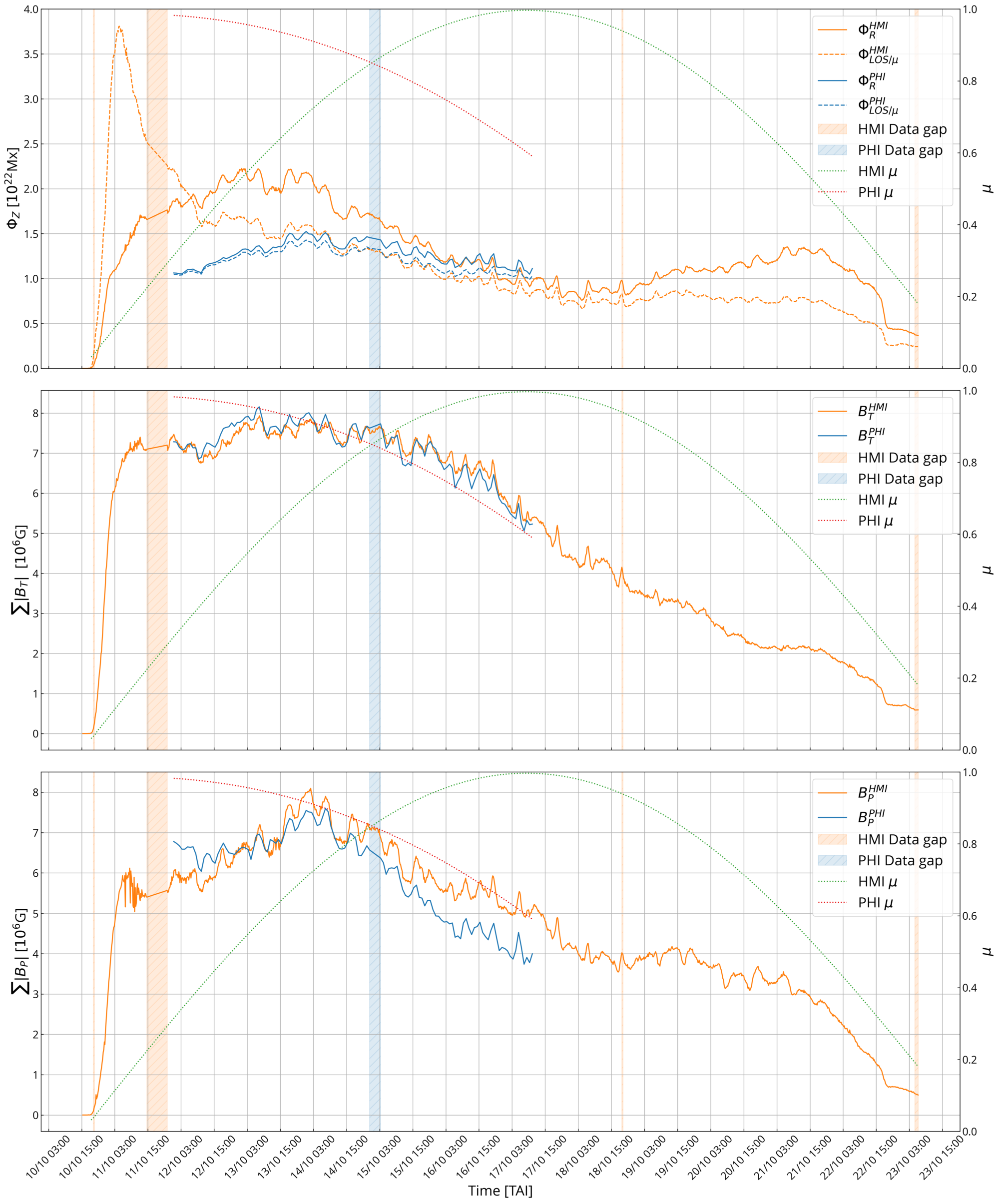}
    {\caption{Unsigned radial magnetic flux and sum of unsigned northward, $B_T$, and westward, $B_P$, magnetic field components during the full duration of the NOAA 13465 passage across the Earth-side disc. All quantities were computed using the SDO/HMI SHARP mask to select the pixels. Top row: unsigned radial flux. Middle row: sum over SHARP mask of unsigned northward magnetic component. Bottom row: sum or unsigned westward magnetic component. The $\mu$ value at the centre of the magnetic field maps for both SO/PHI and SDO/HMI are indicated in all panels. Data gaps from both instruments are highlighted with the hatched area in all panels.} \label{fig:full_sharp}}
\end{figure*}

\begin{figure*}
    \centering
    \includegraphics[width=0.95\textwidth]{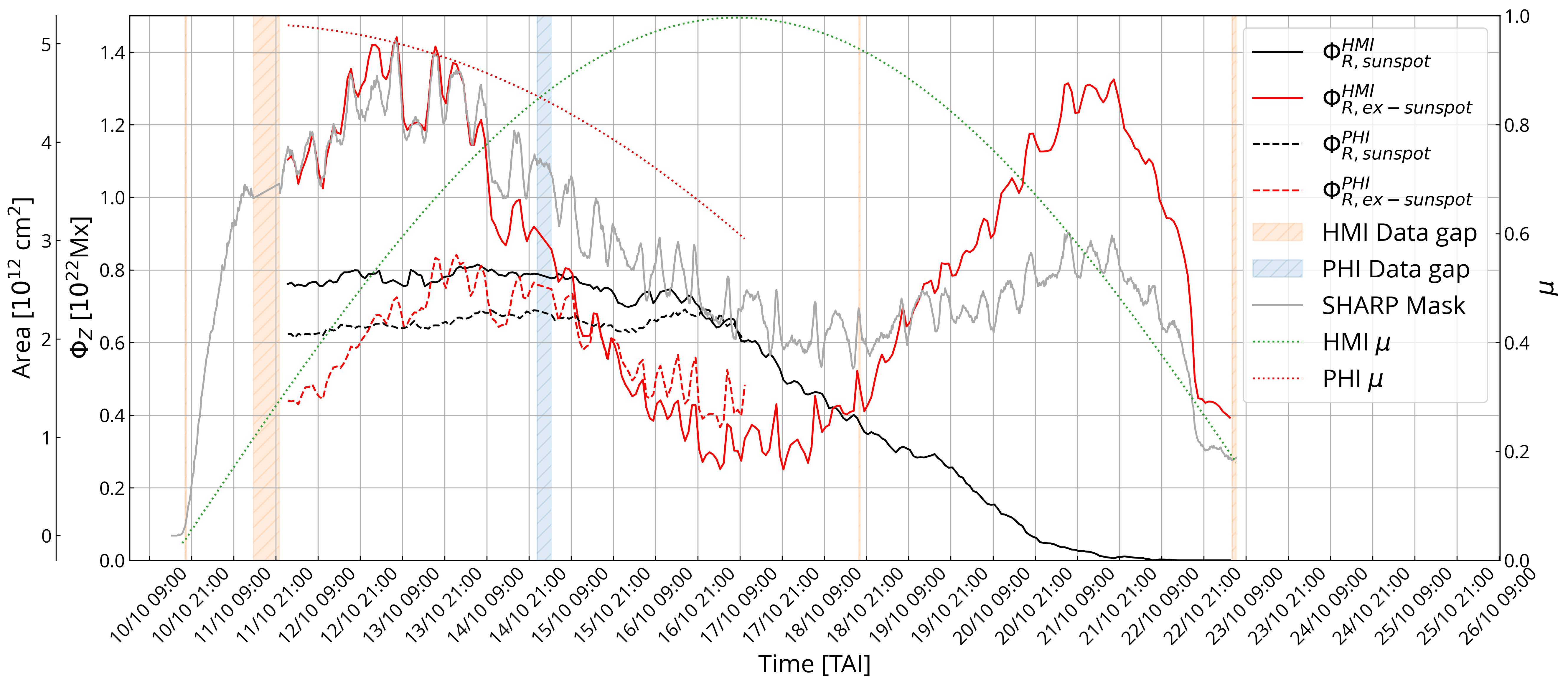}
    {\caption{Unsigned radial magnetic flux outside the sunspot during the full duration of the NOAA 13465 passage across the Earth-side disc. The pixels which constitute the sunspot were defined using a $0.9I_c$ (where $I_c$ is the continuum intensity at disc centre) threshold in the SDO/HMI continuum intensity images from which limb darkening was removed, and then finally applying the SHARP mask. The area of the SHARP mask is indicated in grey. The $\mu$ value at the centre of the magnetic field maps for both SO/PHI and SDO/HMI is given on the axis on the right. Data gaps from both instruments are highlighted by the hatched areas.} \label{fig:sunspot_ex_flux}}
\end{figure*}

\begin{figure*}
    \centering
    \includegraphics[width=\textwidth]{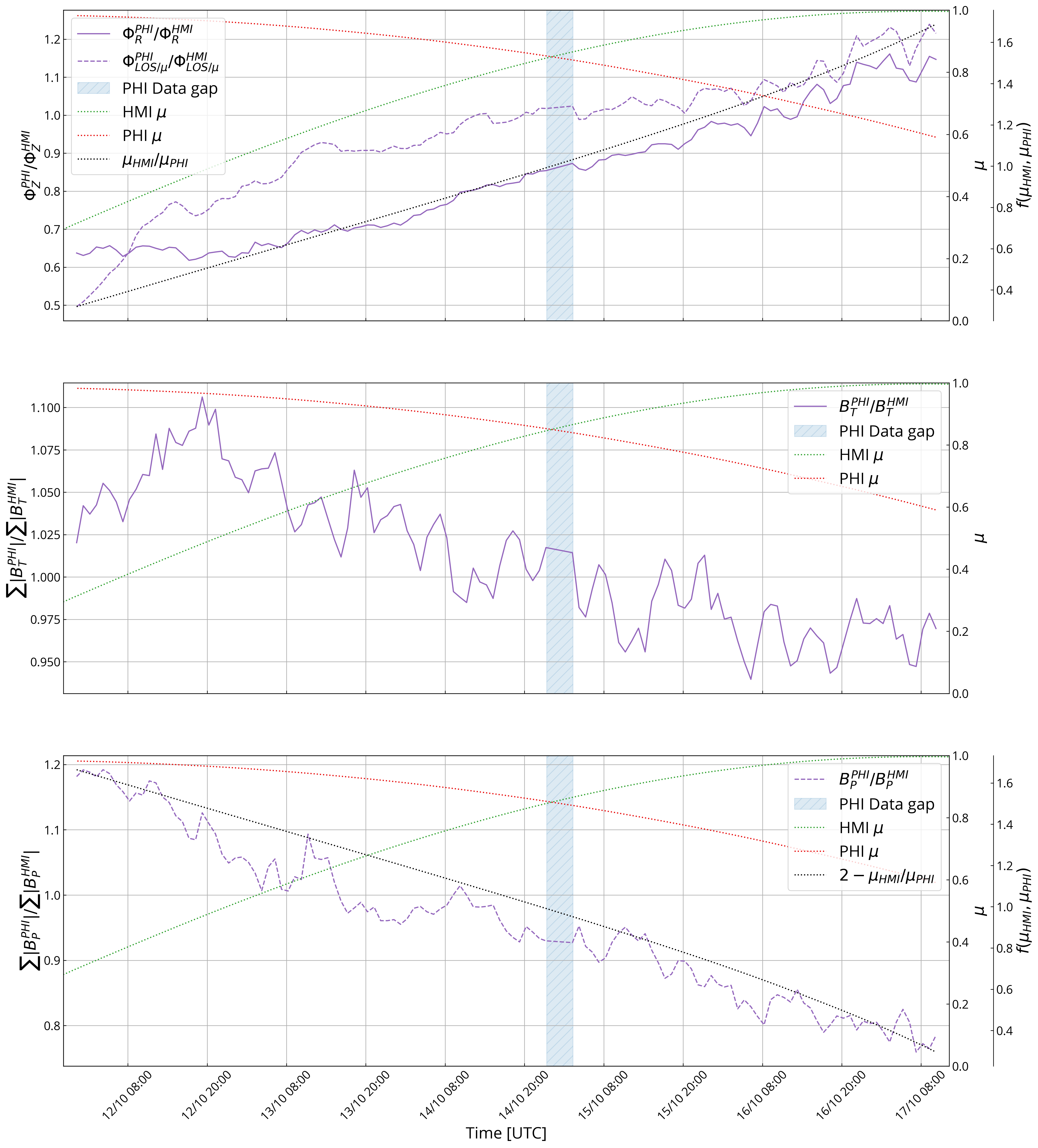}
    {\caption{Ratio of the SO/PHI-HRT and SDO/HMI unsigned vertical magnetic flux and sum of unsigned northward, $B_T$, and westward, $B_P$, magnetic field components during the co-observations of NOAA 13465. All quantities were computed using the SDO/HMI SHARP mask to select the pixels, taking into account the fraction of the SHARP area that SO/PHI-HRT does not initially cover (whitespace in the SO/PHI-HRT images). Top row: unsigned vertical flux, both from the radial component and the $\mu$-corrected line-of-sight magnetic field. Middle row: ratio of the sum of unsigned northward component. Bottom row: ratio of the sum of the unsigned westward magnetic component. The $\mu$ values at the centres of the magnetic field maps for both SO/PHI and SDO/HMI are indicated in all panels by the dotted red (SO/PHI) and green (HMI) lines. In the top panel the ratio of $\mu_{HMI}/\mu_{PHI}$ is denoted by the dotted black line. In addition in the lower panel, the vertical reflection of the $\mu_{HMI}/\mu_{PHI}$ curve at the $\mu_{HMI}=\mu_{PHI}$ line is shown in the dotted black line. Data gaps from both instruments are highlighted with the hatched area in all panels.} \label{fig:full_sharp_ratios}}
\end{figure*}

\begin{figure*}
    \centering
    \includegraphics[width=\textwidth]{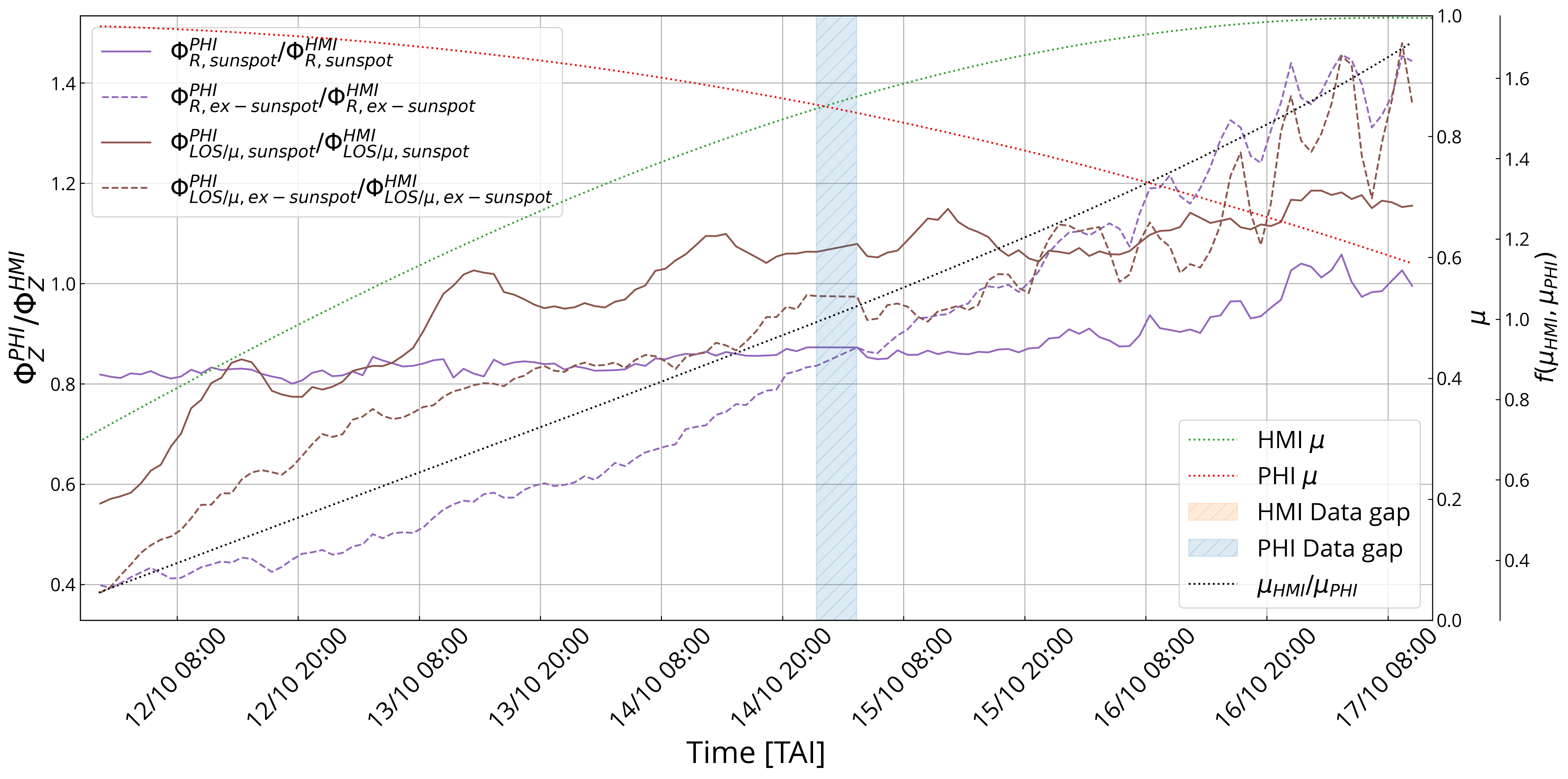}
    {\caption{Ratio of the SO/PHI-HRT and SDO/HMI unsigned vertical magnetic flux in the sunspot, and outside the sunspot. The unsigned vertical flux is derived from both the radial component and the $\mu$-corrected line-of-sight magnetic field. The $\mu$ values at the centres of the magnetic field maps for both SO/PHI and SDO/HMI are indicated in all panels by the dotted red (SO/PHI) and green (HMI) lines. The ratio of $\mu_{HMI}/\mu_{PHI}$ is denoted by the dotted black line. Data gaps from both instruments are highlighted with the hatched area in all panels.} \label{fig:sunspot_plage_ratios}}
\end{figure*}

In the top panel of Fig.~\ref{fig:full_sharp}, the unsigned vertical magnetic flux, $\Phi_{Z}$ from SDO/HMI and SO/PHI-HRT from the full range of observations of NOAA 13465 is shown. For both instruments, the vertical flux is computed using the $B_R$ maps ($\Phi_{R}$), as well as radialised $B_{LOS}/\mu$ maps ($\Phi_{LOS/\mu}$). In all cases, the SDO/HMI SHARP mask is used to select the high confidence pixels to compute the radial flux. For all panels in this figure the quantities are computed using all available pixels that lie within the SDO/HMI SHARP mask. There are a few small areas in this mask that lie outside the SO/PHI-HRT FoV at the start of co-observations. This difference is however considered later when the ratio between the instruments' fluxes are computed. The $\mu$ value at the centre of the instruments' map is highlighted by the dotted red (SO/PHI-HRT) and green (SDO/HMI) lines. Data gaps are indicated by the hatched vertical areas.

The solid orange curve, $\Phi_{R}^{HMI}$, derived from the SDO/HMI $B_{R}$ maps, is characteristic of the SHARP maps, with an initial sharp increase as the active region fully moves into view, and two broad 'shoulders' with peaks at approximate $\mu_{HMI}$ values of 0.6 \citep[see panel A3 of Fig. 6 in][for a similar example]{2014SoPh..289.3549B}. The dotted orange curve, $\Phi_{LOS/\mu}^{HMI}$, derived from the radialised SDO/HMI $B_{LOS}/\mu$ maps, displays a dramatic increase to start, as the sunspot moves across the limb into view, followed by a corresponding sharp decrease until a $\mu_{HMI}$ value of 0.4 when the quantity appears to stabilise. Such a very substantial $B_{LOS}/\mu$ flux excess at low $\mu$ relative to that measured around disc centre was also reported by \cite{2025arXiv251000320A}. Similar large increases in $\Phi_{LOS/\mu}^{HMI}$ near the limb are shown in Fig. 5 of \cite{2025A&A...704A.105K}. To mitigate projection effects, these authors also computed the mean of the $B_{LOS}/\mu$ and $B_{LOS}$ fluxes, as they claim that it represents the temporal evolution better. 

The SO/PHI-HRT $\Phi_{R}^{PHI}$ curve is at a value $40\%$ lower than the corresponding point inferred by SDO/HMI at the same time. As the $\mu_{PHI}$ value decreases, $\Phi^{PHI}_{B_{R}}$ slightly increases, which could also be as a result of the increasing spatial coverage of the region, followed by a slight decrease towards the end of observations. The $\Phi_{R}^{PHI}$ and $\Phi_{R}^{HMI}$ curves only intersect around 16-10-2023 03:00, over 4 days into the co-observations, after which SO/PHI-HRT infers slightly more radial flux.

All curves in all panels show a distinct oscillatory pattern on a 6, 12 and 24 hour periodicity. This arises from the SDO/HMI SHARP mask, and is depicted later in Fig.~\ref{fig:sunspot_ex_flux}, with the selected area oscillating considerably. This is a known systematic of the SDO/HMI data \citep[][]{2014SoPh..289.3483H} due to its orbit. 

At the start, the $\Phi_{LOS/\mu}^{PHI}$ curve is very similar to the $\Phi_{R}^{PHI}$ curve as the active region is very close to disc centre as seen by SO/PHI. As the observations progress the separation of the curves understandably increases. The $\Phi_{LOS/\mu}^{PHI}$ curves from SO/PHI-HRT and SDO/HMI intersect earlier than the $\Phi_{R}^{PHI}$ curves, almost at the point where the $\mu$ values of the two instruments are identical (but from opposite directions), following which the curves separate as the $\Phi_{R}$ curves do, where again the SO/PHI-HRT curve has larger values. 

 As $\Phi_{LOS/\mu}^{HMI}$ is always at a lower value than $\Phi_{R}^{HMI}$, we see here the direct consequence of the underestimate of the strongest fields by the non-vector SDO/HMI $B_{LOS}$ magnetograms compared to those from the vector SDO/HMI magnetograms \citep{2016SoPh..291.1887C}.

In Fig.~\ref{fig:sunspot_ex_flux} the unsigned radial flux inside and outside of the sunspot is displayed from both SDO/HMI and SO/PHI-HRT. In addition, the SHARP mask area is indicated in light grey. The pixels that constitute the sunspot were defined using a $0.9I_c$ (where $I_c$ is the continuum intensity at disc centre) threshold in the SDO/HMI continuum intensity images from which limb darkening was removed, and then finally applying the SHARP mask. During the first 3 days of co-observations the sunspot flux observed by SDO/HMI is stable, following which there is a gradual decrease in $\Phi_{R,sunspot}^{HMI}$ until approximately 02 UTC 17-10-2023, when the flux decreases at a significantly faster rate, signalling the start of the sunspot's decay. Initially $\Phi_{R,sunspot}^{HMI}$ is $20\%$ greater than $\Phi_{R,sunspot}^{PHI}$ but the two converge at 02 UTC 17-10-2023 (as $\Phi_{R,sunspot}^{PHI}$ increased, while $\Phi_{R,sunspot}^{HMI}$ decreased). After this point, the flux from both instruments sharply decreases and are within $<5\%$ of each other. A part of the absolute difference in the sunspot magnetic flux during the initial period of co-observation is attributed to the cross-calibration: SO/PHI-HRT infers $13\%$ lower field strengths in strong field ($B>1.6$\;kG) regions. The sunspot area during the first two days of co-observations is very stable, with a standard deviation of $<3\%$. Early on 14-10-2023 the sunspot area increases $12\%$ before returning to the previous size by the end of 15-10-2023 and remaining stable. From 02 UTC 17-10-2023 the sunspot area massively decreases.

The flux exterior to the sunspot, which we will call ex-sunspot flux in short, displays a much stronger trend. In the SDO/HMI data especially (solid red line), there are two clear shoulders with their maximum at approximately $\mu_{HMI}=0.5$, almost entirely symmetric around $\mu_{HMI}=1$. It is obvious that the flux from features outside the sunspot, which are entirely plage (with the exception of a few pores), exhibits a strong limb effect. Given the exceptional correlation between the SHARP mask area and the ex-sunspot flux, it implies that the SHARP mask itself could be the driving factor behind this limb effect. The ex-sunspot flux observed by SO/PHI-HRT exhibits a similar trend. However the starting point and absolute magnitude of the initial increase of this flux are significantly smaller. This could be because the SHARP mask area initially includes many pixels which SO/PHI-HRT observes to primarily be quiet Sun. Figure~\ref{fig:plage_area} demonstrated that SO/PHI-HRT also infers a greater plage area near to the limb. However, as the SHARP mask towards the end of co-observations uses SDO/HMI data near disc centre, it does not cover all the pixels that SO/PHI-HRT observes as strong plage features, located closer to the limb from its point of view (see lowest row of Fig.~\ref{fig:br_4stage}. This could explain why the ex-sunspot flux from SO/PHI-HRT is only slightly above that of SDO/HMI during this period. 

\subsection{Horizontal magnetic components}

In the middle panel of Fig.~\ref{fig:full_sharp}, the unsigned sum of the northward component from SDO/HMI and SO/PHI-HRT are compared. The two show an excellent agreement throughout, particularly so when the $\mu$ values of both instruments are similar. As SDO and Solar Orbiter during these co-observations were observing at very similar heliographic latitudes (initially $4.3\degree$ separation, reducing to $0.4\degree$ at the end of the co-observations), it stands to reason that both SDO/HMI and SO/PHI-HRT should infer similar magnetic field components that are in the plane perpendicular to both lines-of-sight, and highlights the similarity in their magnetic field inference under similar observational conditions. The ratio of the unsigned northward components, the middle panel of Fig.~\ref{fig:full_sharp_ratios}, oscillates at or near a value of 1, slightly above 1 to start with and slightly below 1 at the end. The mean value is 1.01 with a standard deviation of 0.04.

In the lower panel, the unsigned sum of the westward component is compared. Given that the unsigned quantity is computed, the effect of the east-west bias plays no role here. To start with SO/PHI-HRT infers a larger unsigned westward component, quickly intersecting with that from SDO/HMI after a day of co-observations, after which SDO/HMI infers a larger quantity. The westward magnetic component produces polarisation signals along the line-of-sight when observed near the limb, and given line-of-sight polarisation signals tend to be more reliable, these results imply that in fact the westward component may be overestimated when observing at disc centre. This agrees with results from simulations such as that reported by \cite{2023ApJ...951...23C}.  

\section{Ratio of the vertical magnetic flux and horizontal components}

In the first panel of Fig.~\ref{fig:full_sharp_ratios}, we note a steady increase in $\Phi_{R}^{PHI}/\Phi_{R}^{HMI}$, started at a value between $0.5-0.6$, and increasing beyond $1$ at the end of the co-observations. During the first day there is an initial fast increase in $\Phi_{LOS/\mu}^{PHI}/\Phi_{LOS/\mu}^{HMI}$ followed by a more steady seemingly linear increase afterwards. In contrast, $\Phi_{R}^{PHI}/\Phi_{R}^{HMI}$ is stable around a value of $0.65$ for the first day, corresponds approximately to where $\mu_{HMI}<0.5$, following which a linear increase is found. Remarkably, these linear relationships of the ratios can be expressed as an empirical function of $\mu_{HMI}/\mu_{PHI}$ (dotted black line). For $\Phi_{R}^{PHI}/\Phi_{R}^{HMI}$, we find a Pearson's correlation coefficient of $0.98$ where $\mu_{HMI}>0.5$ and linear regression of:
\begin{equation}
    \Phi^{PHI}_{R}/\Phi^{HMI}_{R} = 0.468 \mu_{HMI}/\mu_{PHI} + 0.369.
\end{equation}
This would imply that at any point where $\mu_{HMI}=\mu_{PHI}$, $\Phi^{PHI}_{R}=0.837\Phi^{HMI}_{R}$. In \cite{sinjan2026_underreview}  the radial magnetic field component between SDO/HMI and SO/PHI-HRT was not compared, so a direct comparison with this result cannot be made.

For $\Phi_{LOS/\mu}^{PHI}/\Phi_{LOS/\mu}^{HMI}$ we find a Pearson's correlation coefficient of $0.96$ and linear regression of:
\begin{equation}
    \Phi^{PHI}_{LOS/\mu}/\Phi^{HMI}_{LOS/\mu} = 0.319 \mu_{HMI}/\mu_{PHI} + 0.669.
\end{equation}
This would imply that at any point where $\mu_{HMI}=\mu_{PHI}$, $\Phi^{PHI}_{LOS/\mu}=0.988\Phi^{HMI}_{LOS/\mu}$. \cite{sinjan2026_underreview} reported a regression fit of $B_{LOS}^{PHI} = 1.1 B_{LOS}^{HMI}+1$\;G, when comparing magnetic fields from SO/PHI-HRT and SDO/HMI at almost the same viewing angle. This fit was primarily affected by the strongest fields in the sunspot, where $B_{LOS}^{PHI}$ was $16\%$ greater than $B_{LOS}^{HMI}$. For the weaker fields, the two instrument's $B_{LOS}$ were almost 1:1. Given the majority of the flux here arises from outside the sunspot, the near equivalence of the radialised line-of-sight magnetic flux at the same viewing angle we find here is in line with the findings in \cite{sinjan2026_underreview}.

The distinctly different behaviour that the ratio of the fluxes derived from the radial and radialised line-of-sight magnetic field exhibit below $\mu=0.5$ is to be explored in future work.

The ratio of the unsigned westward component depicted in the lower panel of Fig.~\ref{fig:full_sharp_ratios}, also demonstrates a clear trend with a steady decrease over time. Similar to the first panel, here we plot the vertical reflection of $\mu_{HMI}/\mu_{PHI}$ around the line $\mu_{HMI}/\mu_{PHI}=1$. This can be expressed as a $f(\mu_{HMI},\mu_{PHI})=2-\mu_{HMI}/\mu_{PHI}$. This function has a clear correlation with the observed ratio. A linear regression using this function provides this expression, with a Pearson's correlation coefficient of $0.97$ and linear regression of:
\begin{equation}
\begin{aligned}
    \sum |B^{PHI}_{P}|/\sum |B^{HMI}_{P}| &= 0.293 (2-\mu_{HMI}/\mu_{PHI}) + 0.652 \\
                                          &= - 0.293 \mu_{HMI}/\mu_{PHI} + 1.238.
\end{aligned}
\end{equation}
This implies, when $\mu_{HMI}=\mu_{PHI}$,$\sum |B^{PHI}_{P}| = 0.945 \sum |B^{HMI}_{P}|$. Similarly to the radial magnetic field case, in \cite{sinjan2026_underreview} the heliographic horizontal magnetic components from SO/PHI-HRT and SDO/HMI were not compared. Nevertheless, the factor of $0.945$ indicates a close quantitative agreement, which is in line with the findings from \cite{sinjan2026_underreview} when comparing the vector magnetic field components in the line-of-sight spherical coordinate system.

In Fig.~\ref{fig:sunspot_plage_ratios}, the ratio of the unsigned vertical flux in the sunspot and outside, namely the plage, is illustrated. The ratio is shown for both the flux derived from the radial and radialised line-of-sight components. 

For the first four days, the ratio $\Phi_{R,sunspot}^{PHI}/\Phi_{R,sunspot}^{HMI}$ is stable around a value of $0.84\pm0.02$. This factor is the same as that implied by the ratio of the total unsigned radial magnetic flux discussed earlier, when $\mu_{HMI}=\mu_{PHI}$. After that point, the ratio increases, as the flux from PHI begins to increase, while that of HMI decreases. This occurs until the sunspot starts to decay at 2 UT 17-10-2023, when the flux ratio is close to unity. The ratio of the flux outside the sunspot, $\Phi_{R,ex-sunspot}^{PHI}/\Phi_{R,ex-sunspot}^{HMI}$ displays a consistent increase throughout however. A linear regression (Pearson correlation coefficient of $0.99$) with $\mu_{HMI}/\mu_{PHI}$ results in:
\begin{equation}
    \Phi^{PHI}_{R,ex-sunspot}/\Phi^{HMI}_{R,ex-sunspot} = 0.838 \mu_{HMI}/\mu_{PHI} + 0.025.
\end{equation}
This demonstrates that the change in the ratio of the total radial magnetic flux is almost exclusively due to the features outside the sunspot, i.e. the magnetic plage. Where $\mu_{HMI}=\mu_{PHI}$, $\Phi^{PHI}_{R,ex-sunspot}=0.863 \Phi^{HMI}_{R,ex-sunspot}$, slightly larger but similar to the factor between of the sunspot and total unsigned flux between the two instruments.

The ratio using the radialised line-of-sight flux displays different characteristics. In the sunspot the ratio increases at first, with a clear periodic trend originating from the orbital systematics from SDO/HMI. Outside the sunspot, the ratio displays a similar trend to that from the radial component, with an absence of orbit induced systematics. The linear regression (Pearson correlation coefficient of $0.97$) with the ratio of the viewing angles is:
\begin{equation}
    \Phi^{PHI}_{LOS/\mu,ex-sunspot}/\Phi^{HMI}_{LOS/\mu,ex-sunspot} = 0.592 \mu_{HMI}/\mu_{PHI} + 0.338,
\end{equation}
implying that $\Phi^{PHI}_{LOS/\mu,ex-sunspot}=0.93\Phi^{HMI}_{LOS/\mu,ex-sunspot}$ when $\mu_{HMI}=\mu_{PHI}$.



\section{Discussion}\label{sec:disc}

One key advantage of the extra vantage point provided by SO/PHI-HRT, is that limb effects can be isolated from solar evolution. This has allowed us to directly investigate any magnetic flux dependencies on the viewing angle, without making any assumptions (such as assuming a given magnetic feature undergoes minimal evolution during its passage across the solar disk). 

The analysis of the plage flux ratios indicates that both the radial and radialised line-of-sight magnetic fluxes are highly correlated with the ratio of the viewing (cosine) angles. However for the sunspot, the radial magnetic flux ratio appeared to be independent until the sunspot began to decay, while the radialised line-of-sight magnetic flux ratio did show a dependence. This confirms that vector magnetograms, instead of line-of-sight magnetograms, are more robust and therefore more suited when studying the evolution of sunspots, or as boundary conditions for magnetic field extrapolations. 



As shown in Sect.~\ref{ssec:br}, the area of the magnetic features is the dominating factor in the change in the plage flux with the viewing angle. In the case of SDO/HMI, the plage flux exhibits a $4-5$ factor increase at $\mu=0.5$ compared to disc centre. The plage magnetic flux inferred by SO/PHI-HRT does not show the same large increase towards the limb. This is, at least partly, because the SHARP mask (deduced from SDO/HMI data) is used to select the pixels from both sets of observations. The SHARP mask has a larger area at the start of co-observations when SO/PHI-HRT observes near disk centre because that was when SDO/HMI was observing near the limb, while towards the end of co-observations, SDO/HMI observes the AR near disk centre, while for SO/PHI-HRT it is near the limb, and therefore the area is not overestimated. As displayed in Fig.~\ref{fig:plage_area}, when the plage area is instead defined using an arbitrary magnetic field strength, both SDO/HMI and SO/PHI-HRT show a dramatic increase in the area near the limb. The consistent behaviour in the plage area dependence on the viewing angle therefore implies that if another method is used to select the `active' pixels in the solar scene, the magnetic flux in SO/PHI-HRT would also show an increase away from disc centre. We cannot rule out however whether the inferred magnetic field has significant limb effects, as was reported by \cite{2023ApJ...951...23C}, as a different approach is needed to isolate this dependency. A full investigation is required into the pixel selection method, which is out of scope for this paper, including comparisons with alternatives such as applying a threshold on the total polarisation in each pixel. 

Some second order processes that may play a role in inducing limb effects include the fact that the instrument observing near the limb samples higher layers in the solar atmosphere, so the measured fields are affected by the expansion of the magnetic field with height and the canopies it forms. Additionally the use of type of inversion employed by both SO/PHI and SDO/HMI may also contribute, in both cases using a simple Milne-Eddington atmosphere. Different inversion setups, single and double atmosphere components were explored in \cite{2026ApJ...999..223Y} and found significant changes in the resulting magnetic fields in a polar observation from Hinode-SOT/SP. 

The co-observations here are somewhat limited by the lowest spatial resolution of the two instruments, in this case that of SDO/HMI. Hence future studies using for example SO/PHI-HRT and Hinode-SOT/SP \citep[Spectropolarimeter on the Solar Optical Telescope,][]{2008SoPh..249..167T,2007SoPh..243....3K} data, which are both obtained at a higher resolution than that of SDO/HMI (as long as Solar Orbiter was not too far from its perihelion), may shed more light on these effects. 

\section{Summary}\label{sec:sum}

In this work we have presented co-observations of active region 13465 with a unique combination of viewing angles from two vantage points. These two vantage points had nearly the same heliographic latitude but different and varying longitudes. The spatial coverage of SO/PHI-HRT almost entirely encompassed this active region, and agreed well with the area demarcated by the SDO/HMI SHARP CEA (cylindrical equal area projection) data series. These co-observations demonstrate considerable differences in the LoS magnetic field maps, with stark contrasts in their polarity and spatial extent of certain features. 

We computed the radial and horizontal photospheric magnetic field components in the heliographic plane in CEA coordinates, in a near identical procedure to that used for the corresponding SDO/HMI SHARP maps, such that they could be directly compared. The maps are in close agreement, particularly so in the active region's sole sunspot. Nevertheless significant differences in the plage regions were observed, namely the spatial extent in $B_{R}$ maps, and polarity in the westward component. The apparent spatial extent of the radial magnetic features increased as either instrument observed closer to the limb. The polarity difference in the westward direction of this plage region is attributed to the East-West hemispheric bias reported in the SDO/HMI data \cite[][]{2022SoPh..297...17L}, which also exists in SO/PHI-FDT data. It remains to be seen if SO/PHI-HRT demonstrates the same bias.

Using the SHARP SDO/HMI mask to select the active region pixels with reliable signals, the radial magnetic flux was compared. With the direct comparison from these two vantage points we were able in part to isolate changes in the flux from solar evolution and study possible systematics when observing near the limb. Analysis of the sunspot and plage magnetic fluxes separately demonstrates that the plage is the dominating contributor to the near limb behaviour of the total unsigned flux. This was predominantly driven by the significant change in apparent area of the plage region, approximately doubling from disc centre to $\mu=0.5$. The higher spatial resolution of SO/PHI-HRT resulted in a smaller relative increase in area towards the limb, hinting that spatial resolution is a key factor in this effect. 

A highly positively correlated linear relationship was found between the ratio of the two instruments' radial and radialised line-of-sight magnetic fluxes, with respect to the ratio of the cosine of the viewing angles. When the ratio was separated into the sunspot and features outside the sunspot, the sunspot radial magnetic flux ratio showed an independence to the viewing angle, suggesting that the radial magnetic field there is more reliable off disc centre. The flux ratio outside the sunspot however showed strong viewing angles dependencies, regardless of whether the radial or radialised line-of-sight magnetic field was used. The unsigned sum of the northward component agreed remarkably well, highlighting the similarity in the magnetic field inference between the two instruments. As this component was almost entirely along the transverse direction for both instruments at all times, so that there is little effect due to the different viewing angles, this further illustrates the concordance of SO/PHI-HRT and SDO/HMI. The unsigned sum of the westward component displayed a significant difference with the viewing angle combinations, with near equivalence when the two instruments observed at the same viewing angle (from opposing directions). The East-West hemispheric bias in SDO/HMI was present in the westward magnetic field, and while it is not yet possible to confirm if the same behaviour exists in SO/PHI-HRT, it most likely is the case, as the full disc telescope of SO/PHI does show the same hemispheric bias.

The derived empirical relationships between the ratio of the two instruments' unsigned vertical magnetic flux and the ratio of their viewing angles, raises a key question regarding the dependence of magnetic flux measurements on the viewing angle. This observed correlation raises the possibility that total solar magnetic flux computations may be considerably affected. Whether this correlation is a true dependence that arises from the finite spatial resolution, area selection method, the magnetic field inference method, or combinations of these, are all to be explored in future studies. These findings must be tested with other co-observations and other instrument pairings. 

With this work we have demonstrated a procedure to reconcile these complex datasets such that they can be directly compared, and first investigated the impact of the magnetic field inference itself, as these observations are used to derive further quantities such as the magnetic free energy and helicity. We have therefore laid the groundwork for future studies, that include whether indeed the magnetic free energy and helicity are too affected by viewing angles. Furthermore the $180\degree$ ambiguity in the magnetic field azimuth is particularly appropriate to be studied with a dataset such as this, so too is the three dimensional magnetic field structure in the higher layers of the solar atmosphere, derived through magnetic field extrapolations using these reconciled datasets as boundary conditions.

Hinode-SOT/SP also co-observed this active region (with almost identical viewing angles as SDO/HMI), providing high quality data to perform a similar analysis together with the SO/PHI-HRT data. Finally, these co-observations are of high relevance to future observations that will be made by the Photospheric Magnetic Imager onboard the Vigil spacecraft, as it will observe the Sun from the Lagrange point L5, i.e. from within the ecliptic plane, but separated longitudinally from Earth by $60$\degree. 

\begin{acknowledgements}
Solar Orbiter is a space mission of international collaboration between ESA and NASA, operated by ESA. We are grateful to the ESA SOC and MOC teams for their support. The German contribution to SO/PHI is funded by the BMWi through DLR and by MPG central funds. The Spanish contribution is funded by AEI/MCIN/10.13039/501100011033/ (RTI2018-096886-C5, PID2021-125325OB-C5, PCI2022-135009-2) and ERDF “A way of making Europe”; “Center of Excellence Severo Ochoa” awards to IAA-CSIC (SEV-2017-0709, CEX2021-001131-S); and a Ramón y Cajal fellowship awarded to DOS. The French contribution is funded by CNES. The SDO/HMI data are courtesy of NASA/SDO and the HMI Science Team. L.A.H is supported by a Royal Society-Research Ireland University Research Fellowship (URF$\backslash$R1$\backslash$241775).
\end{acknowledgements}

\bibliographystyle{bibtex/aa}
\bibliography{biblio}

\begin{thebibliography}{44}
\expandafter\ifx\csname natexlab\endcsname\relax\def\natexlab#1{#1}\fi

\bibitem[{{Albert} {et~al.}(2020){Albert}, {Hirzberger}, {Kolleck}, {Jorge},
  {Busse}, {Rodr{\'\i}guez}, {Carrascosa}, {Fiethe}, {Gandorfer}, {Germerott},
  {Guan}, {Guerrero}, {Gutierrez-Marques}, {Exp{\'o}sito}, {Lange}, {Michalik},
  {Su{\'a}rez}, {Schou}, {Solanki}, {del Toro Iniesta}, \&
  {Woch}}]{2020JATIS...6d8004A}
{Albert}, K., {Hirzberger}, J., {Kolleck}, M., {et~al.} 2020, Journal of
  Astronomical Telescopes, Instruments, and Systems, 6, 048004

\bibitem[{{Albert} {et~al.}(2025){Albert}, {Hirzberger}, {Krivova}, {Li},
  {Calchetti}, {Valori}, {Sinjan}, {Solanki}, {Gandorfer}, {Woch}, {Orozco
  Su{\'a}rez}, \& {Parenti}}]{2025arXiv251000320A}
{Albert}, K., {Hirzberger}, J., {Krivova}, N.~A., {et~al.} 2025, arXiv
  e-prints, arXiv:2510.00320

\bibitem[{{Bail{\'e}n} {et~al.}(2024){Bail{\'e}n}, {Orozco Su{\'a}rez}, {Blanco
  Rodr{\'\i}guez}, {del Toro Iniesta}, {Strecker}, {Moreno Vacas}, {Santamarina
  Guerrero}, {Hirzberger}, {Albert}, {Albelo Jorge}, {Appourchaux},
  {Alvarez-Herrero}, {Gandorfer}, {Germerott}, {Guerrero}, {Gutierrez-Marques},
  {Kahil}, {Kolleck}, {Solanki}, {Volkmer}, {Woch}, {Fiethe}, {G{\'o}mez Cama},
  {P{\'e}rez-Grande}, {Sanchis Kilders}, {Balaguer Jim{\'e}nez}, {Bellot
  Rubio}, {Calchetti}, {Carmona}, {Deutsch}, {Feller}, {Fernandez-Rico},
  {Fern{\'a}ndez-Medina}, {Garc{\'\i}a Parejo}, {Gasent Blesa}, {Gizon},
  {Grauf}, {Heerlein}, {Korpi-Lagg}, {Lange}, {L{\'o}pez Jim{\'e}nez}, {Maue},
  {Meller}, {Michalik}, {M{\"u}ller}, {Nakai}, {Schmidt}, {Schou},
  {Sch{\"u}hle}, {Sinjan}, {Staub}, {Torralbo}, \&
  {Valori}}]{2024A&A...681A..58B}
{Bail{\'e}n}, F.~J., {Orozco Su{\'a}rez}, D., {Blanco Rodr{\'\i}guez}, J.,
  {et~al.} 2024, \aap, 681, A58

\bibitem[{{Bobra} {et~al.}(2014){Bobra}, {Sun}, {Hoeksema}, {Turmon}, {Liu},
  {Hayashi}, {Barnes}, \& {Leka}}]{2014SoPh..289.3549B}
{Bobra}, M.~G., {Sun}, X., {Hoeksema}, J.~T., {et~al.} 2014, \solphys, 289,
  3549

\bibitem[{{Borrero} \& {Kobel}(2011)}]{2011A&A...527A..29B}
{Borrero}, J.~M. \& {Kobel}, P. 2011, \aap, 527, A29

\bibitem[{{Centeno} {et~al.}(2023){Centeno}, {Mili{\'c}}, {Rempel}, {Nitta}, \&
  {Sun}}]{2023ApJ...951...23C}
{Centeno}, R., {Mili{\'c}}, I., {Rempel}, M., {Nitta}, N.~V., \& {Sun}, X.
  2023, \apj, 951, 23

\bibitem[{{Couvidat} {et~al.}(2016){Couvidat}, {Schou}, {Hoeksema}, {Bogart},
  {Bush}, {Duvall}, {Liu}, {Norton}, \& {Scherrer}}]{2016SoPh..291.1887C}
{Couvidat}, S., {Schou}, J., {Hoeksema}, J.~T., {et~al.} 2016, \solphys, 291,
  1887

\bibitem[{{DeForest}(2004)}]{2004SoPh..219....3D}
{DeForest}, C.~E. 2004, \solphys, 219, 3

\bibitem[{{Finley} {et~al.}(2025){Finley}, {Brun}, {Strugarek}, \&
  {Perri}}]{2025A&A...697A.217F}
{Finley}, A.~J., {Brun}, A.~S., {Strugarek}, A., \& {Perri}, B. 2025, \aap,
  697, A217

\bibitem[{{Fletcher} {et~al.}(2011){Fletcher}, {Dennis}, {Hudson}, {Krucker},
  {Phillips}, {Veronig}, {Battaglia}, {Bone}, {Caspi}, {Chen}, {Gallagher},
  {Grigis}, {Ji}, {Liu}, {Milligan}, \& {Temmer}}]{2011SSRv..159...19F}
{Fletcher}, L., {Dennis}, B.~R., {Hudson}, H.~S., {et~al.} 2011, \ssr, 159, 19

\bibitem[{{Gandorfer} {et~al.}(2018){Gandorfer}, {Grauf}, {Staub}, {Bischoff},
  {Woch}, {Hirzberger}, {Solanki}, {{\'A}lvarez-Herrero}, {Garc{\'\i}a Parejo},
  {Schmidt}, {Volkmer}, {Appourchaux}, \& {del Toro
  Iniesta}}]{2018SPIE10698E..4NG}
{Gandorfer}, A., {Grauf}, B., {Staub}, J., {et~al.} 2018, in Society of
  Photo-Optical Instrumentation Engineers (SPIE) Conference Series, Vol. 10698,
  Space Telescopes and Instrumentation 2018: Optical, Infrared, and Millimeter
  Wave, ed. M.~{Lystrup}, H.~A. {MacEwen}, G.~G. {Fazio}, N.~{Batalha},
  N.~{Siegler}, \& E.~C. {Tong}, 106984N

\bibitem[{{Gary} \& {Hagyard}(1990)}]{1990SoPh..126...21G}
{Gary}, G.~A. \& {Hagyard}, M.~J. 1990, \solphys, 126, 21

\bibitem[{{Hoeksema} {et~al.}(2014){Hoeksema}, {Liu}, {Hayashi}, {Sun},
  {Schou}, {Couvidat}, {Norton}, {Bobra}, {Centeno}, {Leka}, {Barnes}, \&
  {Turmon}}]{2014SoPh..289.3483H}
{Hoeksema}, J.~T., {Liu}, Y., {Hayashi}, K., {et~al.} 2014, \solphys, 289, 3483

\bibitem[{{Kahil} {et~al.}(2023){Kahil}, {Gandorfer}, {Hirzberger},
  {Calchetti}, {Sinjan}, {Valori}, {Solanki}, {Van Noort}, {Albert}, {Albelo
  Jorge}, {Alvarez-Herrero}, {Appourchaux}, {Bellot Rubio}, {Blanco
  Rodr{\'\i}guez}, {Feller}, {Fiethe}, {Germerott}, {Gizon}, {Guerrero},
  {Gutierrez-Marques}, {Kolleck}, {Korpi-Lagg}, {Michalik}, {Moreno Vacas},
  {Orozco Su{\'a}rez}, {P{\'e}rez-Grande}, {Sanchis Kilders}, {Schou},
  {Sch{\"u}hle}, {Staub}, {Strecker}, {del Toro Iniesta}, {Volkmer}, \&
  {Woch}}]{2023A&A...675A..61K}
{Kahil}, F., {Gandorfer}, A., {Hirzberger}, J., {et~al.} 2023, \aap, 675, A61

\bibitem[{{Kontogiannis} {et~al.}(2025){Kontogiannis}, {Zhu}, {Barczynski},
  {Stiefel}, {Collier}, {McKevitt}, {Castellanos Dur{\'a}n}, {Berdyugina}, \&
  {Harra}}]{2025A&A...704A.105K}
{Kontogiannis}, I., {Zhu}, Y., {Barczynski}, K., {et~al.} 2025, \aap, 704, A105

\bibitem[{{Kosugi} {et~al.}(2007){Kosugi}, {Matsuzaki}, {Sakao}, {Shimizu},
  {Sone}, {Tachikawa}, {Hashimoto}, {Minesugi}, {Ohnishi}, {Yamada}, {Tsuneta},
  {Hara}, {Ichimoto}, {Suematsu}, {Shimojo}, {Watanabe}, {Shimada}, {Davis},
  {Hill}, {Owens}, {Title}, {Culhane}, {Harra}, {Doschek}, \&
  {Golub}}]{2007SoPh..243....3K}
{Kosugi}, T., {Matsuzaki}, K., {Sakao}, T., {et~al.} 2007, \solphys, 243, 3

\bibitem[{{Krivova} \& {Solanki}(2004)}]{2004A&A...417.1125K}
{Krivova}, N.~A. \& {Solanki}, S.~K. 2004, \aap, 417, 1125

\bibitem[{{Leka} {et~al.}(2009){Leka}, {Barnes}, {Crouch}, {Metcalf}, {Gary},
  {Jing}, \& {Liu}}]{2009SoPh..260...83L}
{Leka}, K.~D., {Barnes}, G., {Crouch}, A.~D., {et~al.} 2009, \solphys, 260, 83

\bibitem[{{Liu} {et~al.}(2022){Liu}, {Gri{\~n}{\'o}n-Mar{\'\i}n}, {Hoeksema},
  {Norton}, \& {Sun}}]{2022SoPh..297...17L}
{Liu}, Y., {Gri{\~n}{\'o}n-Mar{\'\i}n}, A.~B., {Hoeksema}, J.~T., {Norton},
  A.~A., \& {Sun}, X. 2022, \solphys, 297, 17

\bibitem[{{Metcalf}(1994)}]{1994SoPh..155..235M}
{Metcalf}, T.~R. 1994, \solphys, 155, 235

\bibitem[{{Mili{\'c}} {et~al.}(2024){Mili{\'c}}, {Centeno}, {Sun}, {Rempel}, \&
  {de la Cruz Rodr{\'\i}guez}}]{2024A&A...683A.134M}
{Mili{\'c}}, I., {Centeno}, R., {Sun}, X., {Rempel}, M., \& {de la Cruz
  Rodr{\'\i}guez}, J. 2024, \aap, 683, A134

\bibitem[{{M{\"u}ller} {et~al.}(2020){M{\"u}ller}, {St. Cyr}, {Zouganelis},
  {Gilbert}, {Marsden}, {Nieves-Chinchilla}, {Antonucci}, {Auch{\`e}re},
  {Berghmans}, {Horbury}, {Howard}, {Krucker}, {Maksimovic}, {Owen}, {Rochus},
  {Rodriguez-Pacheco}, {Romoli}, {Solanki}, {Bruno}, {Carlsson}, {Fludra},
  {Harra}, {Hassler}, {Livi}, {Louarn}, {Peter}, {Sch{\"u}hle}, {Teriaca}, {del
  Toro Iniesta}, {Wimmer-Schweingruber}, {Marsch}, {Velli}, {De Groof},
  {Walsh}, \& {Williams}}]{2020A&A...642A...1M}
{M{\"u}ller}, D., {St. Cyr}, O.~C., {Zouganelis}, I., {et~al.} 2020, \aap, 642,
  A1

\bibitem[{{Orozco Su{\'a}rez} \& {Del Toro
  Iniesta}(2007)}]{2007A&A...462.1137O}
{Orozco Su{\'a}rez}, D. \& {Del Toro Iniesta}, J.~C. 2007, \aap, 462, 1137

\bibitem[{{Pesnell} {et~al.}(2012){Pesnell}, {Thompson}, \&
  {Chamberlin}}]{2012SoPh..275....3P}
{Pesnell}, W.~D., {Thompson}, B.~J., \& {Chamberlin}, P.~C. 2012, \solphys,
  275, 3

\bibitem[{{Pietarila} {et~al.}(2010){Pietarila}, {Cameron}, \&
  {Solanki}}]{2010A&A...518A..50P}
{Pietarila}, A., {Cameron}, R., \& {Solanki}, S.~K. 2010, \aap, 518, A50

\bibitem[{{Scherrer} {et~al.}(2012){Scherrer}, {Schou}, {Bush}, {Kosovichev},
  {Bogart}, {Hoeksema}, {Liu}, {Duvall}, {Zhao}, {Title}, {Schrijver},
  {Tarbell}, \& {Tomczyk}}]{2012SoPh..275..207S}
{Scherrer}, P.~H., {Schou}, J., {Bush}, R.~I., {et~al.} 2012, \solphys, 275,
  207

\bibitem[{{Schmassmann} {et~al.}(2021){Schmassmann}, {Rempel}, {Bello
  Gonz{\'a}lez}, {Schlichenmaier}, \& {Jur{\v{c}}{\'a}k}}]{2021A&A...656A..92S}
{Schmassmann}, M., {Rempel}, M., {Bello Gonz{\'a}lez}, N., {Schlichenmaier},
  R., \& {Jur{\v{c}}{\'a}k}, J. 2021, \aap, 656, A92

\bibitem[{{Schou} {et~al.}(2012){Schou}, {Scherrer}, {Bush}, {Wachter},
  {Couvidat}, {Rabello-Soares}, {Bogart}, {Hoeksema}, {Liu}, {Duvall}, {Akin},
  {Allard}, {Miles}, {Rairden}, {Shine}, {Tarbell}, {Title}, {Wolfson},
  {Elmore}, {Norton}, \& {Tomczyk}}]{2012SoPh..275..229S}
{Schou}, J., {Scherrer}, P.~H., {Bush}, R.~I., {et~al.} 2012, \solphys, 275,
  229

\bibitem[{{Sinjan} {et~al.}(2023){Sinjan}, {Calchetti}, {Hirzberger}, {Kahil},
  {Valori}, {Solanki}, {Albert}, {Albelo Jorge}, {Alvarez-Herrero},
  {Appourchaux}, {Bellot Rubio}, {Blanco Rodr{\'\i}guez}, {Feller},
  {Gandorfer}, {Germerott}, {Gizon}, {G{\'o}mez Cama}, {Guerrero},
  {Gutierrez-Marques}, {Kolleck}, {Korpi-Lagg}, {Michalik}, {Moreno Vacas},
  {Orozco Su{\'a}rez}, {P{\'e}rez-Grande}, {Sanchis Kilders}, {Balaguer
  Jim{\'e}nez}, {Schou}, {Sch{\"u}hle}, {Staub}, {Strecker}, {del Toro
  Iniesta}, {Volkmer}, \& {Woch}}]{2023A&A...673A..31S}
{Sinjan}, J., {Calchetti}, D., {Hirzberger}, J., {et~al.} 2023, \aap, 673, A31

\bibitem[{{Sinjan} {et~al.}(2022){Sinjan}, {Calchetti}, {Hirzberger}, {Orozco
  Su{\'a}rez}, {Albert}, {Albelo Jorge}, {Appourchaux}, {Alvarez-Herrero},
  {Blanco Rodr{\'\i}guez}, {Gandorfer}, {Germerott}, {Guerrero}, {Gutierrez
  Marquez}, {Kahil}, {Kolleck}, {Solanki}, {del Toro Iniesta}, {Volkmer},
  {Woch}, {Fiethe}, {G{\'o}mez Cama}, {P{\'e}rez-Grande}, {Sanchis Kilders},
  {Balaguer Jim{\'e}nez}, {Bellot Rubio}, {Carmona}, {Deutsch},
  {Fernandez-Rico}, {Fern{\'a}ndez-Medina}, {Garc{\'\i}a Parejo}, {Gasent
  Blesa}, {Gizon}, {Grauf}, {Heerlein}, {Korpi-Lagg}, {Lange}, {L{\'o}pez
  Jim{\'e}nez}, {Maue}, {Meller}, {Michalik}, {Moreno Vacas}, {M{\"u}ller},
  {Nakai}, {Schmidt}, {Schou}, {Sch{\"u}hle}, {Staub}, {Strecker}, {Torralbo},
  \& {Valori}}]{2022SPIE12189E..1JS}
{Sinjan}, J., {Calchetti}, D., {Hirzberger}, J., {et~al.} 2022, in Society of
  Photo-Optical Instrumentation Engineers (SPIE) Conference Series, Vol. 12189,
  Software and Cyberinfrastructure for Astronomy VII, 121891J

\bibitem[{{Sinjan} {et~al.}(\noop{3001}under review 2026){Sinjan},
  {Hirzberger}, {Valori}, {Solanki}, \& {Hayes}}]{sinjan2026_underreview}
{Sinjan}, J., {Hirzberger}, J., {Valori}, G., C.~D., {Solanki}, S.~K., \&
  {Hayes}, L.~A. \noop{3001}under review 2026, \aap

\bibitem[{{Sinjan} {et~al.}(2024){Sinjan}, {Solanki}, {Hirzberger},
  {Riethm{\"u}ller}, \& {Przybylski}}]{2024A&A...690A.341S}
{Sinjan}, J., {Solanki}, S.~K., {Hirzberger}, J., {Riethm{\"u}ller}, T.~L., \&
  {Przybylski}, D. 2024, \aap, 690, A341

\bibitem[{{Solanki}(2003)}]{2003A&ARv..11..153S}
{Solanki}, S.~K. 2003, \aapr, 11, 153

\bibitem[{{Solanki} {et~al.}(2020){Solanki}, {del Toro Iniesta}, {Woch},
  {Gandorfer}, {Hirzberger}, {Alvarez-Herrero}, {Appourchaux}, {Mart{\'\i}nez
  Pillet}, {P{\'e}rez-Grande}, {Sanchis Kilders}, {Schmidt}, {G{\'o}mez Cama},
  {Michalik}, {Deutsch}, {Fernandez-Rico}, {Grauf}, {Gizon}, {Heerlein},
  {Kolleck}, {Lagg}, {Meller}, {M{\"u}ller}, {Sch{\"u}hle}, {Staub}, {Albert},
  {Alvarez Copano}, {Beckmann}, {Bischoff}, {Busse}, {Enge}, {Frahm},
  {Germerott}, {Guerrero}, {L{\"o}ptien}, {Meierdierks}, {Oberdorfer},
  {Papagiannaki}, {Ramanath}, {Schou}, {Werner}, {Yang}, {Zerr}, {Bergmann},
  {Bochmann}, {Heinrichs}, {Meyer}, {Monecke}, {M{\"u}ller}, {Sperling},
  {{\'A}lvarez Garc{\'\i}a}, {Aparicio}, {Balaguer Jim{\'e}nez}, {Bellot
  Rubio}, {Cobos Carracosa}, {Girela}, {Hern{\'a}ndez Exp{\'o}sito}, {Herranz},
  {Labrousse}, {L{\'o}pez Jim{\'e}nez}, {Orozco Su{\'a}rez}, {Ramos},
  {Barandiar{\'a}n}, {Bastide}, {Campuzano}, {Cebollero}, {D{\'a}vila},
  {Fern{\'a}ndez-Medina}, {Garc{\'\i}a Parejo}, {Garranzo-Garc{\'\i}a},
  {Laguna}, {Mart{\'\i}n}, {Navarro}, {N{\'u}{\~n}ez Peral}, {Royo},
  {S{\'a}nchez}, {Silva-L{\'o}pez}, {Vera}, {Villanueva}, {Fourmond}, {de
  Galarreta}, {Bouzit}, {Hervier}, {Le Clec'h}, {Szwec}, {Chaigneau},
  {Buttice}, {Dominguez-Tagle}, {Philippon}, {Boumier}, {Le Cocguen},
  {Baranjuk}, {Bell}, {Berkefeld}, {Baumgartner}, {Heidecke}, {Maue}, {Nakai},
  {Scheiffelen}, {Sigwarth}, {Soltau}, {Volkmer}, {Blanco Rodr{\'\i}guez},
  {Domingo}, {Ferreres Sabater}, {Gasent Blesa}, {Rodr{\'\i}guez
  Mart{\'\i}nez}, {Osorno Caudel}, {Bosch}, {Casas}, {Carmona}, {Herms},
  {Roma}, {Alonso}, {G{\'o}mez-Sanjuan}, {Piqueras}, {Torralbo}, {Fiethe},
  {Guan}, {Lange}, {Michel}, {Bonet}, {Fahmy}, {M{\"u}ller}, \&
  {Zouganelis}}]{2020A&A...642A..11S}
{Solanki}, S.~K., {del Toro Iniesta}, J.~C., {Woch}, J., {et~al.} 2020, \aap,
  642, A11

\bibitem[{{Solanki} {et~al.}(1998){Solanki}, {Steiner}, {Buente}, {Murphy}, \&
  {Ploner}}]{1998A&A...333..721S}
{Solanki}, S.~K., {Steiner}, O., {Buente}, M., {Murphy}, G., \& {Ploner},
  S.~R.~O. 1998, \aap, 333, 721

\bibitem[{{Sun}(2013)}]{2013arXiv1309.2392S}
{Sun}, X. 2013, arXiv e-prints, arXiv:1309.2392

\bibitem[{{SunPy Community} {et~al.}(2020){SunPy Community}, {Barnes}, {Bobra},
  {Christe}, {Freij}, {Hayes}, {Ireland}, {Mumford}, {Perez-Suarez}, {Ryan},
  {Shih}, {Chanda}, {Glogowski}, {Hewett}, {Hughitt}, {Hill}, {Hiware},
  {Inglis}, {Kirk}, {Konge}, {Mason}, {Maloney}, {Murray}, {Panda}, {Park},
  {Pereira}, {Reardon}, {Savage}, {Sip{\H{o}}cz}, {Stansby}, {Jain}, {Taylor},
  {Yadav}, {Rajul}, \& {Dang}}]{2020ApJ...890...68S}
{SunPy Community}, {Barnes}, W.~T., {Bobra}, M.~G., {et~al.} 2020, \apj, 890,
  68

\bibitem[{{Tsuneta} {et~al.}(2008){Tsuneta}, {Ichimoto}, {Katsukawa}, {Nagata},
  {Otsubo}, {Shimizu}, {Suematsu}, {Nakagiri}, {Noguchi}, {Tarbell}, {Title},
  {Shine}, {Rosenberg}, {Hoffmann}, {Jurcevich}, {Kushner}, {Levay}, {Lites},
  {Elmore}, {Matsushita}, {Kawaguchi}, {Saito}, {Mikami}, {Hill}, \&
  {Owens}}]{2008SoPh..249..167T}
{Tsuneta}, S., {Ichimoto}, K., {Katsukawa}, Y., {et~al.} 2008, \solphys, 249,
  167

\bibitem[{{Turmon} {et~al.}(2010){Turmon}, {Jones}, {Malanushenko}, \&
  {Pap}}]{2010SoPh..262..277T}
{Turmon}, M., {Jones}, H.~P., {Malanushenko}, O.~V., \& {Pap}, J.~M. 2010,
  \solphys, 262, 277

\bibitem[{{van Driel-Gesztelyi} \& {Green}(2015)}]{2015LRSP...12....1V}
{van Driel-Gesztelyi}, L. \& {Green}, L.~M. 2015, Living Reviews in Solar
  Physics, 12, 1

\bibitem[{{Volkmer} {et~al.}(2012){Volkmer}, {Bosch}, {Feger}, {Gomez},
  {Heidecke}, {Schmidt}, {Scheiffelen}, {Sigwarth}, \&
  {Soltau}}]{2012SPIE.8442E..4PV}
{Volkmer}, R., {Bosch}, J., {Feger}, B., {et~al.} 2012, in Society of
  Photo-Optical Instrumentation Engineers (SPIE) Conference Series, Vol. 8442,
  Space Telescopes and Instrumentation 2012: Optical, Infrared, and Millimeter
  Wave, ed. M.~C. {Clampin}, G.~G. {Fazio}, H.~A. {MacEwen}, \& J.~M.
  {Oschmann}, Jr., 84424P

\bibitem[{{Webb} \& {Howard}(2012)}]{2012LRSP....9....3W}
{Webb}, D.~F. \& {Howard}, T.~A. 2012, Living Reviews in Solar Physics, 9, 3

\bibitem[{{Yamashiro} {et~al.}(2026){Yamashiro}, {Sun}, {Mili{\'c}}, {Quintero
  Noda}, {Pastor Yabar}, {Centeno}, {Liu}, {Go{\v{s}}i{\'c}}, \&
  {Yang}}]{2026ApJ...999..223Y}
{Yamashiro}, B., {Sun}, X., {Mili{\'c}}, I., {et~al.} 2026, \apj, 999, 223

\bibitem[{{Zouganelis} {et~al.}(2020){Zouganelis}, {De Groof}, {Walsh},
  {Williams}, {M{\"u}ller}, {St Cyr}, {Auch{\`e}re}, {Berghmans}, {Fludra},
  {Horbury}, {Howard}, {Krucker}, {Maksimovic}, {Owen},
  {Rodr{\'\i}guez-Pacheco}, {Romoli}, {Solanki}, {Watson}, {Sanchez}, {Lefort},
  {Osuna}, {Gilbert}, {Nieves-Chinchilla}, {Abbo}, {Alexandrova},
  {Anastasiadis}, {Andretta}, {Antonucci}, {Appourchaux}, {Aran}, {Arge},
  {Aulanier}, {Baker}, {Bale}, {Battaglia}, {Bellot Rubio}, {Bemporad},
  {Berthomier}, {Bocchialini}, {Bonnin}, {Brun}, {Bruno}, {Buchlin},
  {B{\"u}chner}, {Bucik}, {Carcaboso}, {Carr}, {Carrasco-Bl{\'a}zquez},
  {Cecconi}, {Cernuda Cangas}, {Chen}, {Chitta}, {Chust}, {Dalmasse},
  {D'Amicis}, {Da Deppo}, {De Marco}, {Dolei}, {Dolla}, {Dudok de Wit}, {van
  Driel-Gesztelyi}, {Eastwood}, {Espinosa Lara}, {Etesi}, {Fedorov},
  {F{\'e}lix-Redondo}, {Fineschi}, {Fleck}, {Fontaine}, {Fox}, {Gandorfer},
  {G{\'e}not}, {Georgoulis}, {Gissot}, {Giunta}, {Gizon}, {G{\'o}mez-Herrero},
  {Gontikakis}, {Graham}, {Green}, {Grundy}, {Haberreiter}, {Harra}, {Hassler},
  {Hirzberger}, {Ho}, {Hurford}, {Innes}, {Issautier}, {James}, {Janitzek},
  {Janvier}, {Jeffrey}, {Jenkins}, {Khotyaintsev}, {Klein}, {Kontar},
  {Kontogiannis}, {Krafft}, {Krasnoselskikh}, {Kretzschmar}, {Labrosse},
  {Lagg}, {Landini}, {Lavraud}, {Leon}, {Lepri}, {Lewis}, {Liewer}, {Linker},
  {Livi}, {Long}, {Louarn}, {Malandraki}, {Maloney}, {Martinez-Pillet},
  {Martinovic}, {Masson}, {Matthews}, {Matteini}, {Meyer-Vernet}, {Moraitis},
  {Morton}, {Musset}, {Nicolaou}, {Nindos}, {O'Brien}, {Orozco Suarez},
  {Owens}, {Pancrazzi}, {Papaioannou}, {Parenti}, {Pariat}, {Patsourakos},
  {Perrone}, {Peter}, {Pinto}, {Plainaki}, {Plettemeier}, {Plunkett}, {Raines},
  {Raouafi}, {Reid}, {Retino}, {Rezeau}, {Rochus}, {Rodriguez},
  {Rodriguez-Garcia}, {Roth}, {Rouillard}, {Sahraoui}, {Sasso}, {Schou},
  {Sch{\"u}hle}, {Sorriso-Valvo}, {Soucek}, {Spadaro}, {Stangalini}, {Stansby},
  {Steller}, {Strugarek}, {{\v{S}}tver{\'a}k}, {Susino}, {Telloni}, {Terasa},
  {Teriaca}, {Toledo-Redondo}, {del Toro Iniesta}, {Tsiropoula}, {Tsounis},
  {Tziotziou}, {Valentini}, {Vaivads}, {Vecchio}, {Velli}, {Verbeeck},
  {Verdini}, {Verscharen}, {Vilmer}, {Vourlidas}, {Wicks},
  {Wimmer-Schweingruber}, {Wiegelmann}, {Young}, \&
  {Zhukov}}]{2020A&A...642A...3Z}
{Zouganelis}, I., {De Groof}, A., {Walsh}, A.~P., {et~al.} 2020, \aap, 642, A3

\end{thebibliography}

\begin{appendix}

\onecolumn
\section{SO/PHI-FDT Low Latency data}\label{append:fdtll}

\begin{figure*}[ht!]
    \centering
    \includegraphics[width=0.8\textwidth]{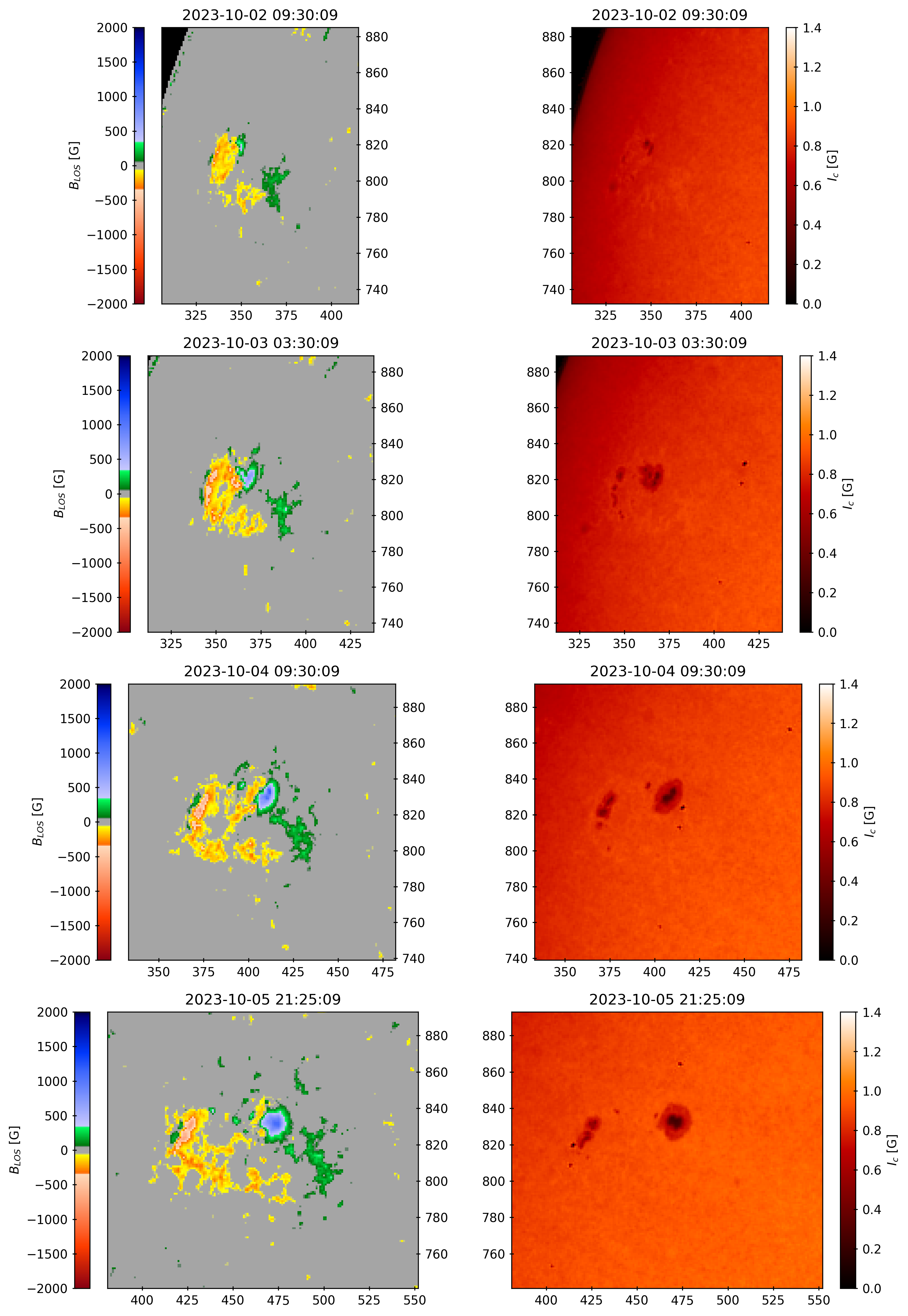}
    {\caption{Line-of-sight magnetic field and continuum intensity from SO/PHI-FDT low latency datasets of NOAA AR 13465 on the far side of the Sun. Values below $\pm200$ G are masked as the low latency datasets are produced with a simplified data reduction process.} \label{fig:fdt_ll_1}}
\end{figure*}

\begin{figure*}[ht!]
    \centering
    \includegraphics[width=0.8\textwidth]{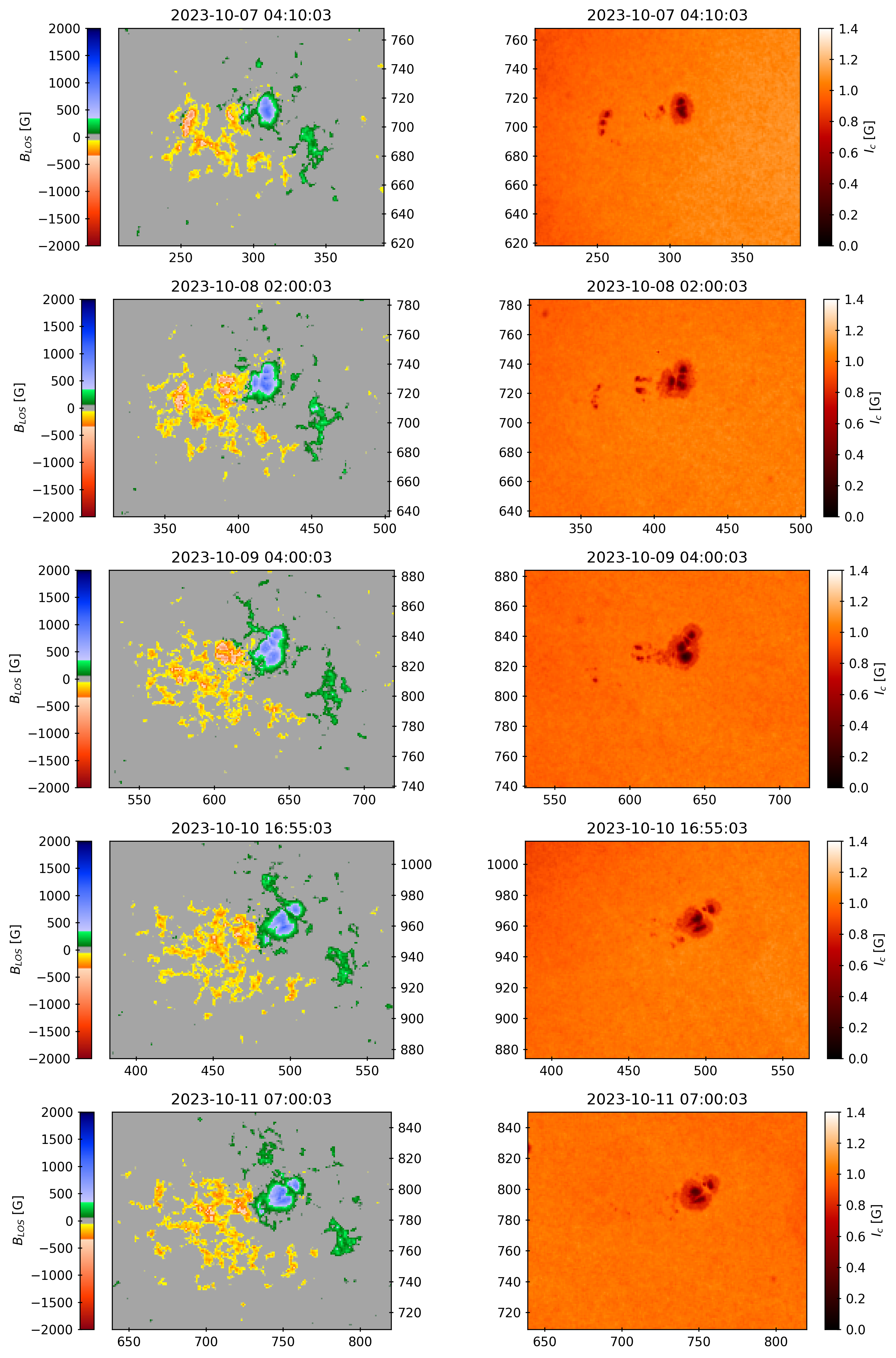}
    {\caption{Line-of-sight magnetic field and continuum intensity from SO/PHI-FDT Low Latency datasets of NOAA AR 13465 on the far side of the Sun. Values below $\pm200$ G are masked as the low latency datasets are produced with a simplified data reduction process.} \label{fig:fdt_ll_2}}
\end{figure*}

\section{SO/PHI-HRT and SDO/HMI Line-of-sight magnetic field maps}\label{append:blos}

\begin{figure*}[ht!]
    \centering
    \includegraphics[width=0.87\textwidth]{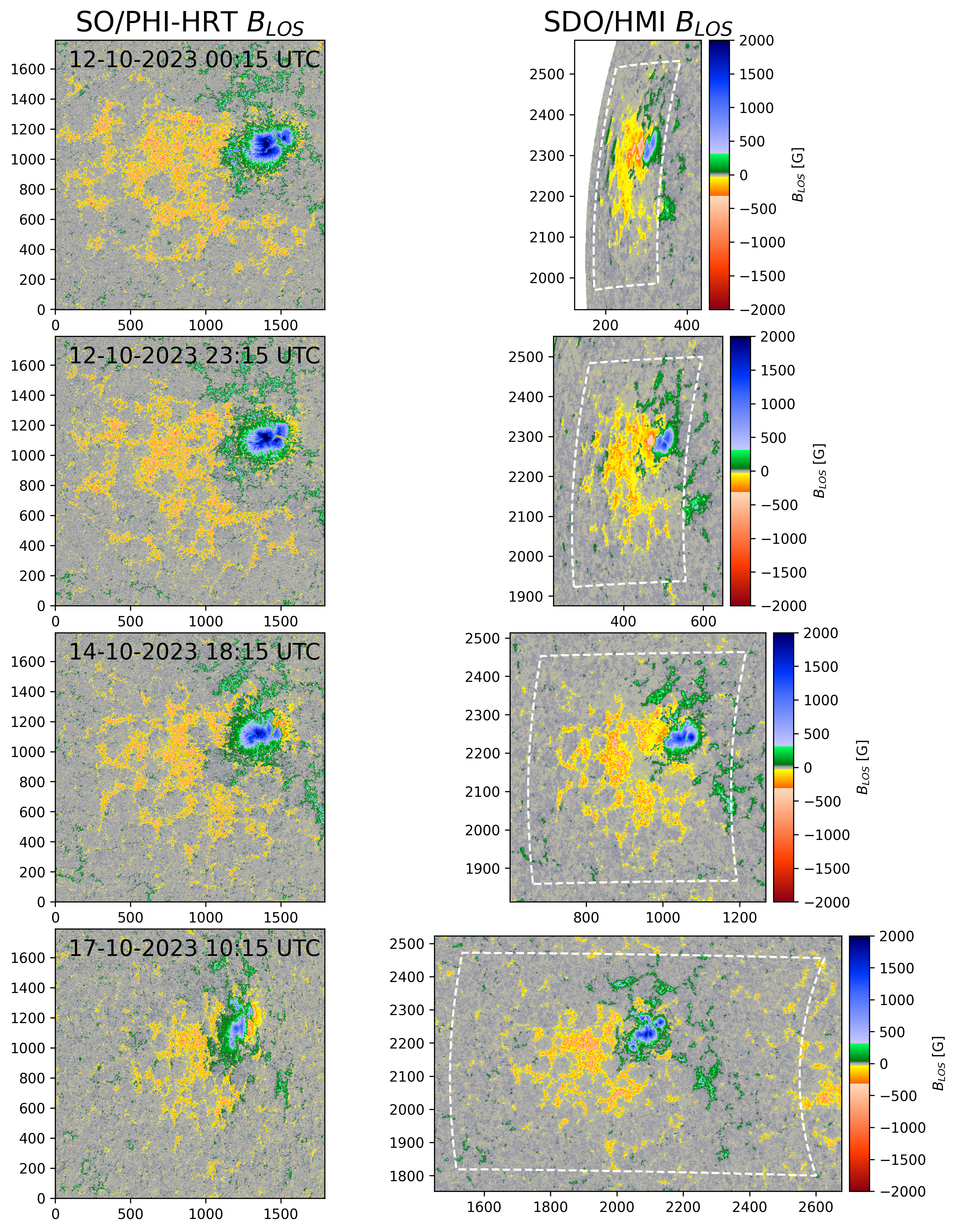}
    {\caption{Line-of-sight magnetic field of the active region as co-observed by SO/PHI-HRT and SDO/HMI. The four rows are observations at the same times as the rows in Fig.~\ref{fig:icnt1}. Note the difference in spatial resolution and spatial extent. The time of SO/PHI-HRT observations is shown in the text overlay with the nearest $720$s observation from SDO/HMI plotted beside the corresponding SO/PHI-HRT map. The FoV of SO/PHI-HRT is indicated by the dashed white contours in the SDO/HMI cutout.} \label{fig:blos1}}
\end{figure*}

\newpage

\section{SDO/HMI SHARP Disambiguation}

\begin{figure*}[ht!]
    \centering
    \includegraphics[width=0.95\textwidth]{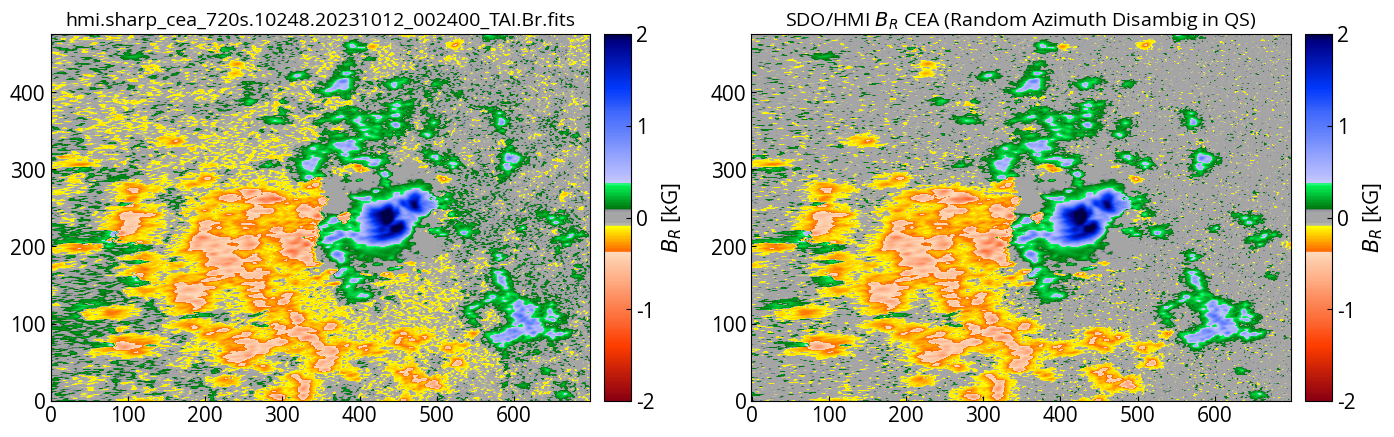}
    {\caption{Left: SDO/HMI $B_R$ at 12-10-2023 00:24 TAI as downloaded from JSOC. Right: manually computed SDO/HMI $B_R$ map on the same SHARP CEA grid using the random solution for disambiguation of the azimuth in the quiet Sun (QS) regions.} \label{fig:hmi_disambig_example}}
\end{figure*}

\end{appendix}

\end{document}